\documentclass[twocolumn]{aastex63}

\newcommand\tess{TESS}
\newcommand\gaia{\textit{Gaia}}

\newcommand{\unit}[1]{\ensuremath{\, \mathrm{#1}}} 

\usepackage[T1]{fontenc} 
\usepackage[utf8]{inputenc} 

\submitjournal{ApJ}

\begin{document}

\title{GJ 3929: High Precision Photometric and Doppler Characterization of an Exo-Venus and its Hot, Mini-Neptune-mass Companion}


\author[0000-0001-7708-2364]{Corey Beard}
\affiliation{Department of Physics and Astronomy, 4129 Frederick Reines Hall, University of California, Irvine, Irvine, CA, 92697, USA}


\author[0000-0003-0149-9678]{Paul Robertson}
\affil{Department of Physics and Astronomy, 4129 Frederick Reines Hall, University of California, Irvine, Irvine, CA, 92697, USA}

\author[0000-0001-8401-4300]{Shubham Kanodia}
\affil{Department of Astronomy \& Astrophysics, 525 Davey Laboratory, The Pennsylvania State University, University Park, PA, 16802, USA}
\affil{Center for Exoplanets and Habitable Worlds, 525 Davey Laboratory, The Pennsylvania State University, University Park, PA, 16802, USA}
\affil{Penn State Extraterrestrial Intelligence Center, 525 Davey Laboratory, The Pennsylvania State University, University Park, PA, 16802, USA}

\author[0000-0001-8342-7736]{Jack Lubin}
\affil{Department of Physics and Astronomy, 4129 Frederick Reines Hall, University of California, Irvine, Irvine, CA, 92697, USA}


\author[0000-0003-4835-0619]{Caleb I. Ca\~nas}
\affil{Department of Astronomy \& Astrophysics, 525 Davey Laboratory, The Pennsylvania State University, University Park, PA, 16802, USA}
\affil{Center for Exoplanets and Habitable Worlds, 525 Davey Laboratory, The Pennsylvania State University, University Park, PA, 16802, USA}

\author[0000-0002-5463-9980]{Arvind F.\ Gupta}
\affil{Department of Astronomy \& Astrophysics, 525 Davey Laboratory, The Pennsylvania State University, University Park, PA, 16802, USA}
\affil{Center for Exoplanets and Habitable Worlds, 525 Davey Laboratory, The Pennsylvania State University, University Park, PA, 16802, USA}

\author[0000-0002-5034-9476]{Rae Holcomb}
\affil{Department of Physics and Astronomy, 4129 Frederick Reines Hall, University of California, Irvine, Irvine, CA, 92697, USA}

\author[0000-0002-7227-2334]{Sinclaire Jones}
\affiliation{Princeton University, Department of Astrophysical Sciences, 4 Ivy Lane, Princeton, NJ 08540, USA}

\author[0000-0002-2990-7613]{Jessica E.~Libby-Roberts}
\affil{Department of Astronomy \& Astrophysics, 525 Davey Laboratory, The Pennsylvania State University, University Park, PA, 16802, USA}
\affil{Center for Exoplanets and Habitable Worlds, 525 Davey Laboratory, The Pennsylvania State University, University Park, PA, 16802, USA}

\author[0000-0002-9082-6337]{Andrea S.J.\ Lin}
\affil{Department of Astronomy \& Astrophysics, 525 Davey Laboratory, The Pennsylvania State University, University Park, PA, 16802, USA}
\affil{Center for Exoplanets and Habitable Worlds, 525 Davey Laboratory, The Pennsylvania State University, University Park, PA, 16802, USA}

\author[0000-0001-9596-7983]{Suvrath Mahadevan}
\affil{Department of Astronomy \& Astrophysics, 525 Davey Laboratory, The Pennsylvania State University, University Park, PA, 16802, USA}
\affil{Center for Exoplanets and Habitable Worlds, 525 Davey Laboratory, The Pennsylvania State University, University Park, PA, 16802, USA}


\author[0000-0001-7409-5688]{Guðmundur Stefánsson}
\affil{Henry Norris Russell Fellow}
\affil{Department of Astrophysical Sciences, Princeton University, 4 Ivy Lane, Princeton, NJ 08540, USA}



\author[0000-0003-4384-7220]{Chad F.\ Bender}
\affil{Steward Observatory, The University of Arizona, 933 N.\ Cherry Ave, Tucson, AZ 85721, USA}

\author[0000-0002-6096-1749]{Cullen H.\ Blake}
\affil{Department of Physics and Astronomy, University of Pennsylvania, 209 S 33rd St, Philadelphia, PA 19104, USA}

\author[0000-0001-9662-3496]{William D. Cochran}
\affiliation{McDonald Observatory and Center for Planetary Systems Habitability, The University of Texas, Austin TX 78712 USA.}


\author[0000-0002-7714-6310]{Michael Endl}
\affiliation{Center for Planetary Systems Habitability, The University of Texas at Austin, Austin, TX 78712, USA}
\affiliation{Department of Astronomy, The University of Texas at Austin, TX, 78712, USA}


\author[0000-0002-0885-7215]{Mark Everett}
\affiliation{National Optical Infrared Astronomy Research Laboratory 950 N. Cherry Ave., Tucson, AZ 85719}

\author[0000-0001-6545-639X]{Eric B.\ Ford}
\affil{Department of Astronomy \& Astrophysics, 525 Davey Laboratory, The Pennsylvania State University, University Park, PA, 16802, USA}
\affil{Center for Exoplanets and Habitable Worlds, 525 Davey Laboratory, The Pennsylvania State University, University Park, PA, 16802, USA}
\affil{Institute for Computational and Data Sciences, The Pennsylvania State University, University Park, PA, 16802, USA}
\affil{Center for Astrostatistics, 525 Davey Laboratory, The Pennsylvania State University, University Park, PA, 16802, USA}

\author[0000-0002-0560-1433]{Connor Fredrick}
\affil{Time and Frequency Division, National Institute of Standards and Technology, 325 Broadway, Boulder, CO 80305, USA}
\affil{Department of Physics, University of Colorado, 2000 Colorado Avenue, Boulder, CO 80309, USA}


\author[0000-0003-1312-9391]{Samuel Halverson}
\affil{Jet Propulsion Laboratory, California Institute of Technology, 4800 Oak Grove Drive, Pasadena, California 91109}


\author[0000-0003-1263-8637]{Leslie Hebb}
\affiliation{Department of Physics, Hobart and William Smith Colleges, 300 Pulteney Street, Geneva,
NY, 14456, USA}


\author[0000-0001-7318-6318]{Dan Li}
\affil{NSF's National Optical-Infrared Astronomy Research Laboratory, 950 N.\ Cherry Ave., Tucson, AZ 85719, USA}

\author[0000-0002-9632-9382]{Sarah E.\ Logsdon}
\affil{NSF's National Optical-Infrared Astronomy Research Laboratory, 950 N.\ Cherry Ave., Tucson, AZ 85719, USA}


\author[0000-0002-4927-9925]{Jacob Luhn}
\affil{Department of Physics and Astronomy, 4129 Frederick Reines Hall, University of California, Irvine, Irvine, CA, 92697, USA}

\author[0000-0003-0241-8956]{Michael W.\ McElwain}
\affil{Exoplanets and Stellar Astrophysics Laboratory, NASA Goddard Space Flight Center, Greenbelt, MD 20771, USA}

\author[0000-0001-5000-1018]{Andrew J. Metcalf}
\affiliation{Space Vehicles Directorate, Air Force Research Laboratory, 3550 Aberdeen Ave. SE, Kirtland AFB, NM 87117, USA}
\affiliation{Time and Frequency Division, National Institute of Technology, 325 Broadway, Boulder, CO 80305, USA} 
\affiliation{Department of Physics, 390 UCB, University of Colorado Boulder, Boulder, CO 80309, USA}


\author[0000-0001-8720-5612]{Joe P.\ Ninan}
\affil{Department of Astronomy and Astrophysics, Tata Institute of Fundamental Research, Homi Bhabha Road, Colaba, Mumbai 400005, India}


\author[0000-0002-2488-7123]{Jayadev Rajagopal}
\affil{NSF's National Optical-Infrared Astronomy Research Laboratory, 950 N.\ Cherry Ave., Tucson, AZ 85719, USA}


\author[0000-0001-8127-5775]{Arpita Roy}
\affil{Space Telescope Science Institute, 3700 San Martin Dr, Baltimore, MD 21218, USA}
\affil{Department of Physics and Astronomy, Johns Hopkins University, 3400 N Charles St, Baltimore, MD 21218, USA}

\author[0000-0003-2435-130X]{Maria Schutte}
\affiliation{Homer L. Dodge Department of Physics and Astronomy, University of Oklahoma, 440 W. Brooks Street, Norman, OK 73019, USA}

\author[0000-0002-4046-987X]{Christian Schwab}
\affil{Department of Physics and Astronomy, Macquarie University, Balaclava Road, North Ryde, NSW 2109, Australia }

\author[0000-0002-4788-8858]{Ryan C. Terrien}
\affil{Carleton College, One North College St., Northfield, MN 55057, USA}

\author[0000-0001-9209-1808]{John Wisniewski}
\affiliation{Homer L. Dodge Department of Physics and Astronomy, University of Oklahoma, 440 W. Brooks Street, Norman, OK 73019, USA}

\author[0000-0001-6160-5888]{Jason T.\ Wright}
\affil{Department of Astronomy \& Astrophysics, 525 Davey Laboratory, The Pennsylvania State University, University Park, PA, 16802, USA}
\affil{Center for Exoplanets and Habitable Worlds, 525 Davey Laboratory, The Pennsylvania State University, University Park, PA, 16802, USA}
\affil{Penn State Extraterrestrial Intelligence Center, 525 Davey Laboratory, The Pennsylvania State University, University Park, PA, 16802, USA}

\correspondingauthor{Corey Beard}
\email{ccbeard@uci.edu}

\begin{abstract}
We detail the follow up and characterization of a transiting exo-Venus identified by TESS, GJ 3929b, (TOI-2013b) and its non-transiting companion planet, GJ 3929c (TOI-2013c). GJ 3929b is an Earth-sized exoplanet in its star's Venus-zone (P$_{b}$ = 2.616272 $\pm$ 0.000005 days; S$_{b}$ = 17.3$^{+0.8}_{-0.7}$ S$_{\oplus}$) orbiting a nearby M dwarf. GJ 3929c is most likely a non-transiting sub-Neptune. Using the new, ultra-precise NEID spectrometer on the WIYN 3.5\,m Telescope at Kitt Peak National Observatory, we are able to modify the mass constraints of planet b reported in previous works and consequently improve the significance of the mass measurement to almost 4$\sigma$ confidence (M$_{b}$ = 1.75 $\pm$ 0.45 M$_{\oplus}$). We further adjust the orbital period of planet c from its alias at 14.30 $\pm$ 0.03 days to the likely true period of 15.04 $\pm$ 0.03 days, and adjust its minimum mass to m$\sin i$ = 5.71 $\pm$ 0.92 M$_{\oplus}$. Using the diffuser-assisted ARCTIC imager on the ARC 3.5 m telescope at Apache Point Observatory, in addition to publicly available TESS and LCOGT photometry, we are able to constrain the radius of planet b to R$_{p}$ = 1.09 $\pm$ 0.04 R$_{\oplus}$. GJ 3929b is a top candidate for transmission spectroscopy in its size regime (TSM = 14 $\pm$ 4), and future atmospheric studies of GJ 3929b stand to shed light on the nature of small planets orbiting M dwarfs.
\end{abstract}


\section{Introduction} \label{sec:intro}

Transit photometry has become an extremely important technique for the characterization of exoplanets, and has long been the most fruitful method for identifying candidates in the first place \citep{borucki10}. With the advent of the Transiting Exoplanet Survey Satellite \citep[TESS;][]{ricker15}, we are in a new era of exoplanet detection around low-mass stars. Since the beginning of the TESS mission, over 5000 new exoplanet candidates have been discovered. Many identified candidates are false positives, however, and observations using different techniques are often required to accurately characterize orbital periods, rule out false positive scenarios, detect longer period or non-transiting companions, or to measure additional parameters of an exoplanet \citep[e.g.,][]{weiss21,kanodia21,lubin22, canas22}. Radial velocity (RV) observations are a particularly important follow-up method, as they allow for 1) independent confirmation of a transiting planet signal, 2) characterization of a planet's mass, and 3) a search for non-transiting companion planets.

A transiting exoplanet candidate with a 2.6 day period was first identified by the TESS Science Processing and Operations Center \citep[SPOC; ][]{jenkins16} pipeline around the nearby (15.822 $\pm$ 0.006) M dwarf GJ 3929 on 2020 June 19, then designated TOI-2013.01. Our team began follow-up observations using RV instruments and high contrast imaging shortly after this announcement, with the intent to confirm the planetary nature of this system and refine the planetary parameters of the transiting candidate.

\cite{kemmer22} recently published an analysis of the system, placing constraints on its planetary parameters and validating its planetary nature, as well as discovering a second, non-transiting planet candidate during their analysis. \cite{kemmer22} were unable to precisely constrain the mass of the transiting planet (K/$\sigma$ = 2.88), however, possibly due to the unanticipated existence of planet c.

Using precise RVs obtained with the NEID spectrograph on the WIYN\footnote{The WIYN Observatory is a joint facility of the University of Wisconsin–Madison, Indiana University, NSF’s NOIRLab, the Pennsylvania State University, Purdue University, and the University of California, Irvine.} 3.5 m telescope at Kitt Peak National Observatory, RVs taken with the Habitable Zone Planet Finder, and previously published CARMENES RV data, we were able to refine the mass measurements of both planets in the system. Using two ground-based transits obtained with the diffuser-assisted ARCTIC imager, in addition to publicly available TESS and LCOGT data, we refine the radius measurement of this system. Furthermore, our analysis concludes that the additional non-transiting planet candidate has a period of $\sim$ 15 days, and we upgrade GJ 3929c from a candidate to a planet.

In Section \ref{sec:observations}, we give a summary of the data used in our analysis. In Section \ref{sec:stellar}, we detail our estimation of the system's stellar parameters. In Section \ref{sec:analysis}, we detail the steps taken to measure planetary and orbital parameters, and the investigation of an additional planet. In Section \ref{sec:discussion}, we discuss our findings, and the implications for the system. Finally, Section \ref{sec:summary} summarizes our results and conclusions.

\section{Observations}\label{sec:observations}

A summary of our observational data and key properties is visible in Table \ref{tab:obstable}.

\begin{deluxetable*}{lcccc}
\tablecaption{Summary of Observational Data. \label{tab:obstable}}
\tablehead{\colhead{Instrument}&  \colhead{Date Range}&
\colhead{RMS}& \colhead{Average Error}&
\colhead{Type}}
\startdata
\multicolumn{4}{l}{\hspace{-0.2cm}} \\
TESS & 2020 April 16 - 2020 June 8 & 1346 ppm & 1441 ppm& Photometry \\
ARCTIC & 2021 February 27 - 2021 April 30 & 1000 ppm & 734 ppm& Photometry \\
LCO & 2021 April 15 & 1522 ppm & 692 ppm & Photometry \\
CARMENES & 2020 July 30 - 2021 July 19 & 3.87 m s$^{-1}$ & 1.97 m s$^{-1}$& RV \\
HPF & 2021 August 27 - 2022 March 11 & 8.81 m s$^{-1}$ & 8.42 m s$^{-1}$& RV \\
NEID & 2021 January 6 - 2022 January 27 & 10.6 m s$^{-1}$ & 1.55 m s$^{-1}$& RV \\
\enddata
\end{deluxetable*}

\subsection{TESS}\label{sec:TESS}

\begin{figure*}[] 
\centering
\includegraphics[width=\textwidth]{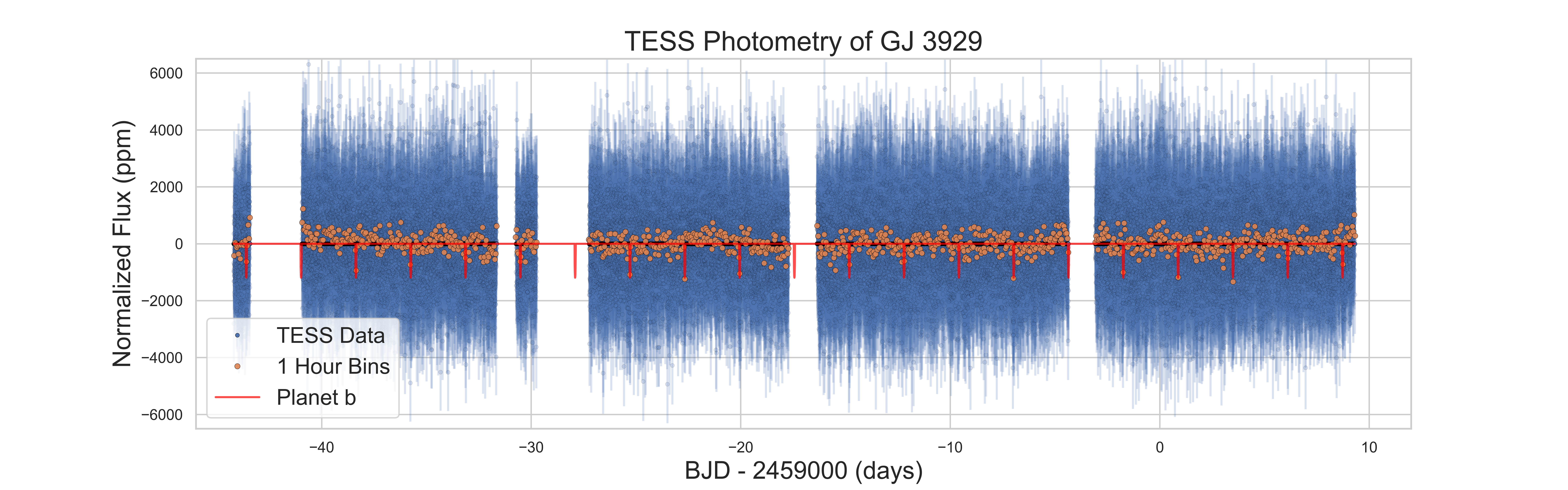}
\caption{PDCSAP flux of GJ 3929 as taken during TESS Sectors 24 and 25. Overlaid is data binned into one hour intervals. Additionally, we plot a maximum a-posteriori (MAP) fit of the transits of planet b. A phase fold of the transits after our complete analysis is visible in Figure \ref{fig:transit_final}. The transit model is described in Section \ref{sec:transitanalysis}.} \label{fig:TESS}
\end{figure*}

GJ 3929 was observed by the TESS spacecraft between 2020 April 16 and 2020 June 8. These dates correspond to Sectors 24 and 25 of the TESS nominal mission. GJ 3929 was observed in CCD 1 of Camera 1 during sector 24, and CCD 2 of Camera 1 during sector 25. The TESS photometry were first reduced by SPOC. After initial processing, we used the pre-search data conditioning simple aperture photometry \citep[PDCSAP;][]{stumpe12} in our analysis. Data points flagged as poor quality are discarded before analysis. A plot of the TESS PDCSAP flux used in the analysis is shown in Figure \ref{fig:TESS}.

\subsection{Ground based photometric follow up}\label{sec:photometry}

Ground-based follow-up can be a useful tool not only to validate the planetary nature of transiting signals, but to refine the measured parameters of transiting exoplanets. Here we detail the ground-based photometric follow-up of for GJ 3929b.

\subsubsection{ARCTIC}\label{sec:arctic}

We observed three transits of GJ 3929b on the nights of 2021 February 26, 2021 April 30, and 2021 September 21, using the Astrophysical Research Consortium (ARC) Telescope Imaging Camera \citep[ARCTIC;][]{huehnerhoff_astrophysical_2016} at the ARC 3.5 m Telescope at Apache Point Observatory (APO). To achieve precise photometry on nearby bright stars, we used the engineered diffuser described in \citet{stefansson_toward_2017}.

The airmass of GJ 3929 varied from 1.00 to 1.66 over the course of its observation on 2021 February 26. The observations were performed using a 30 nm wide narrowband Semrock filter centered at 857 nm \citep[described in ][]{stefansson18} due to moderate cloud coverage, with an exposure time of  33.1$\unit{s}$ in the quad-readout mode with $2\times2$ on-chip binning. In the $2\times2$ binning mode, ARCTIC has a gain of 2\unit{e/ADU}, a plate scale of 0.228 \unit{\arcsec/pixel}, and a readout time of 2.7 $\unit{s}$. We reduced the raw data using \texttt{AstroImageJ} \citep{collins17}. We selected a photometric aperture of 31 pixels (7.07\unit{\arcsec}) and used an annulus with an inner radius of 70 pixels (15.96\unit{\arcsec}), and an outer radius of 100 pixels (22.8\unit{\arcsec}).

We also observed a transit of GJ 3929b on 2021 April 30. The airmass during observations varied between 1.00 and 1.51. The observations were performed using the same Semrock filter as described previously, with an exposure time of  45$\unit{s}$ in the quad-readout mode with $2\times2$ on-chip binning. For the final reduction, we selected a photometric aperture of 33 pixels (7.52\unit{\arcsec}) and used an annulus with an inner radius of 58 pixels (13.22\unit{\arcsec}), and an outer radius of 87 pixels (19.84\unit{\arcsec}).

We observed a final transit of GJ 3929b on 2021 September 21. The airmass during observations varied between 1.21 and 3.22, and the resulting scatter in data points was $>$ 3 times the values of either previous ARCTIC night (rms$_{20210226}$ = 1000 ppm; rms$_{20210430}$ = 910 ppm; rms$_{20210921}$ = 3400 ppm). Consequently, we chose not to use this final ARCTIC transit during analysis of planet b.

We checked for airmass correlation on each night, but found little evidence for any significant correlation. A plot of the ARCTIC transits used in our final analysis is visible in Figure \ref{fig:transit_final}.

\subsubsection{LCOGT}

We additionally use publicly available data taken by the Las Cumbres Observatory Global Telescope Network \citep[LCOGT; ][]{brown13}. These data were obtained from the Exoplanet Follow-up Observing Program (ExoFOP) website \footnote{https://exofop.ipac.caltech.edu/tess/}. Two transits of GJ 3929b were obtained using the LCOGT. The first transit was obtained on 2021 April 10. Data were taken by both the SINISTRO CCDs at the 1 m telescopes of the McDonald Observatory (McD) and the Cerro Tololo Interamerican Observatory (CTIO). Both instruments have a pixel scale of 0.00389 pix$^{-1}$ and a FOV of 260 × 260.

A second transit was obtained on 2021 April 15. These data were taken simultaneously in 4 different filters (g$^{\prime}$, i$^{\prime}$, r$^{\prime}$, and z$_{s}^{\prime}$) with the Multi-color Simultaneous Camera for studying Atmospheres of Transiting exoplanets 3 camera \citep[MuSCAT3;][]{narita20} mounted on the 2 m Faulkes Telescope North at Haleakala Observatory (HAL). It has a pixel scale of 0.0027 pix$^{-1}$ corresponding to a FOV of 9.01×9.01. 

As outlined in \cite{kemmer22}, high airmass caused the CTIO observations to exhibit higher scatter. In fact, both transits on 2021 April 10 exhibit much higher scatter (rms$_{CTIO}$ = 3300 ppm; rms$_{MCD}$ = 2200 ppm) than on 2021 April 15 (rms$_{gp}$ = 1010 ppm; rms$_{ip}$ = 850 ppm; rms$_{rp}$ = 910 ppm; rms$_{zs}$ = 920 ppm). Consequently, for the same reasons outlined in Section \ref{sec:arctic}, we chose not to utilize either transit from 2021 April 10 in our final analysis.

The publicly available data were calibrated by the LCOGT BANZAI pipeline \citep{mccully18}, and photometric data were extracted using AstroImageJ \citep{collins17}. The resulting photometric data are the same that were utilized in \cite{kemmer22}.

\subsection{High Contrast Imaging}

High contrast imaging can be important for ruling out false positive scenarios. \cite{kemmer22} used high-resolution images obtained from the AstraLux camera \citep{hormuth08} at the Calar Alto Observatory to rule out false positive scenarios. They were able to rule out nearby luminous sources down to a $\Delta$z$^{\prime}$ $<$ 5.5 at 1$\arcsec$. Here we detail our team's adaptive optics (AO) follow up of GJ 3929b, and add to the evidence of a planetary explanation for the transit events.

\subsubsection{ShARCS on the Shane telescope}\label{sec:shane}

We observed GJ 3929 using the ShARCS camera on the Shane 3 m telescope at Lick Observatory \citep{srinath14}. GJ 3929 was observed using the $K_S$ and $J$ filters on the night of 2021 February 26. Instrument repairs prevented our observations from benefiting from Laser Guide Star (LGS) mode. Fortunately, GJ 3929 is sufficiently bright such that LGS mode is helpful, but not necessary. Further instrument repairs prevented our observations from using a dither-routine to create master-sky images of GJ 3929. Instead, after a series of observations, we shifted several arcseconds to an empty region of sky, and took images with the same exposure time for purposes of sky subtraction.

The raw data are reduced using a custom pipeline developed by our team \citep[described in][]{beard22}. Using algorithms from \cite{espinoza16}, we then generate a 5 sigma contrast curve as the final part of our analysis. We detect no companions at a $\Delta$K$_s$= 4.85 at 0.76$\arcsec$ and $\Delta$Ks = 9.75 at 8.35$\arcsec$. Additionally, we detect no companions at $\Delta$J = 4.54 at 1.09$\arcsec$ and $\Delta$J = 7.62 at 8.99$\arcsec$.

\begin{figure}[] 
\centering
\includegraphics[width=0.5\textwidth]{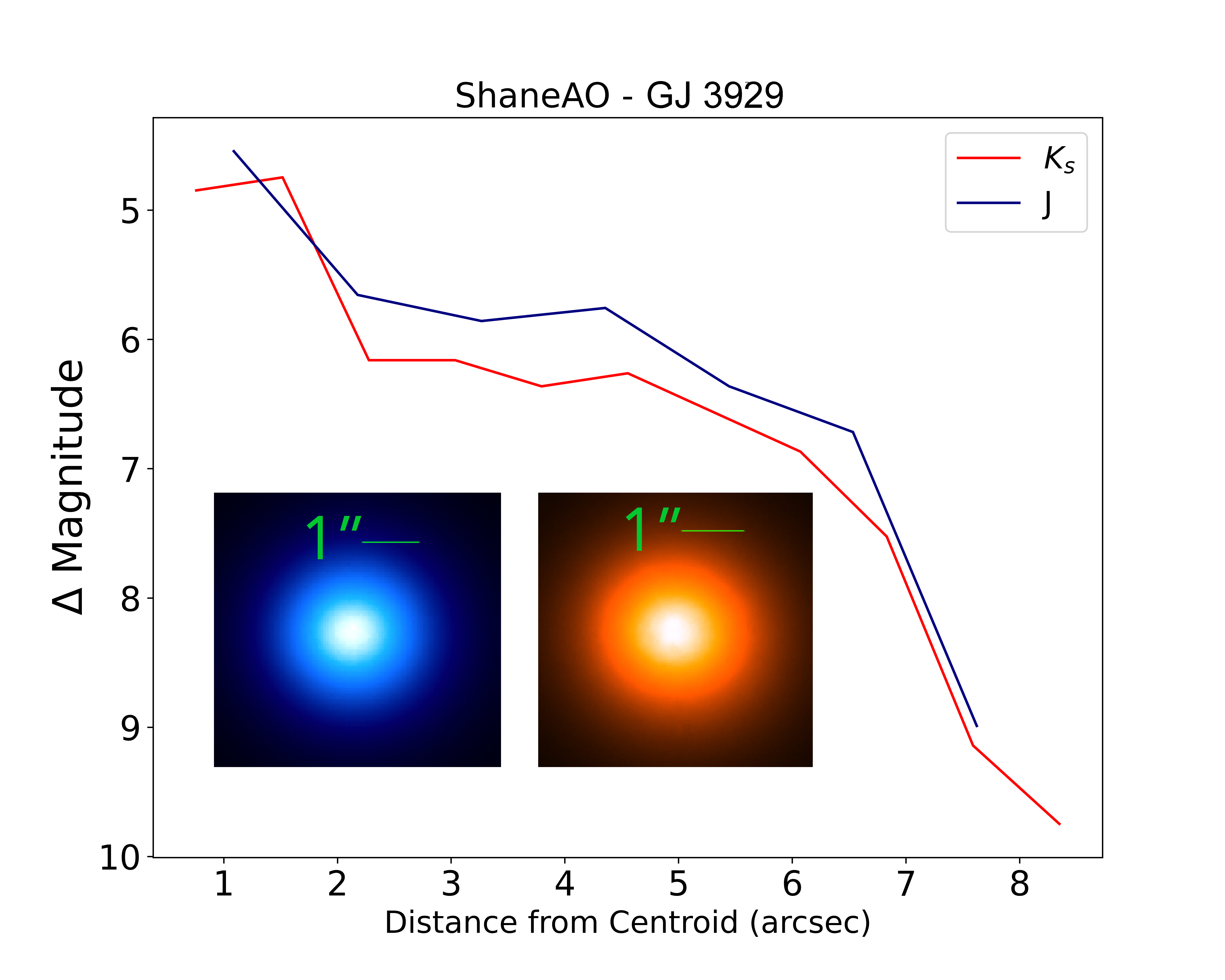}
\caption{5$\sigma$ contrast curves of GJ 3929 taken using the K$_{s}$ and J filters. The data were taken on 2021 February 26. The overcast conditions and poor seeing on 2021 February 26 resulted in challenges with sky-subtraction. As a result, the magnitude difference between the centroid and background don't drop as quickly as expected.} \label{fig:shane}
\end{figure}

We note that observing conditions on 26 February 2021 were marginal. As a result of overcast conditions and poor seeing, the FWHM of the centroid in each reduced AO image was fairly large (0.77\unit{\arcsec} and 1.11\unit{\arcsec}). Consequently, our final constraints on nearby luminous companions are not as tight as they might have been. However, our high contrast images were taken in redder wavebands than than the z$^{\prime}$ filter used in \cite{kemmer22}, and so we provide additional sensitivity toward detecting redder, cooler companions. In tandem, our results and those outlined in \cite{kemmer22} are consistent: we detect no nearby luminous companions as an explanation for the observed transit event.

\subsection{Radial Velocity Follow-Up}\label{sec:hpfrvs}

We obtained RVs of GJ 3929b in order to constrain the mass of the system and to independently confirm the planetary nature of the transiting planet. Here we detail the RV data acquired for the system GJ 3929.

\subsubsection{The NEID Spectrometer on the WIYN 3.5 m Telescope at KPNO}\label{sec:neidobservations}

We obtained RVs of GJ 3929 using the new, ultra-precise NEID spectrometer \citep{schwab16} on the WIYN 3.5 m telescope at Kitt Peak National Observatory (KPNO). NEID is an environmentally stabilized \citep{stefansson16,robertson19} fiber-fed spectrograph \citep{kanodia18a} with broad wavelength coverage (3800 - 9300 \AA). We observed GJ 3929 in High Resolution (HR) mode with an average resolving power $\mathcal{R}$ = 110,000. The default NEID pipeline utilizes the Cross-Correlation Function \citep[CCF; ][]{baranne96} method to produce RVs. However, this method tends to be less effective on M dwarfs \citep[e.g.,][]{anglada-escude12}, and so we use a modified version of the \texttt{SpEctrum Radial Velocity AnaLyser} pipeline \citep[\texttt{SERVAL};][]{zechmeister18} as described in \cite{stefansson21}. \texttt{SERVAL} shifts and combines all observed spectra into a master template, and compares this template with known reference spectra. We then minimize the \(\chi^2\) statistic to determine the shifts of each observed spectrum. We mask telluric and sky-emission lines during this process. A telluric mask is calculated based on their predicted locations using \texttt{telfit} \citep{gullikson14}, a Python wrapper to the Line-by-Line Radiative Transfer Model package \citep{clough05}.

We obtained 27 observations of GJ 3929 between 2021 January 6 and 2022 January 27. Our first two nights of observation for this system used 3 consecutive 900 s exposures, but we later changed our observation strategy to one 1800 s exposure per night. We obtained a median SNR of 44.8 in order 102 ($\lambda$ = 4942 \AA) of NEID for each unbinned observation. The median unbinned RV errorbar is 1.18 m s$^{-1}$. The errorbars are estimated from expected photon noise. A total of 23 nightly binned RVs were obtained, though 4 were discarded because the laser frequency comb calibrator was not available on those nights, resulting in a less precise instrument drift solution that is insufficient for precision RV analysis. This left us with 19 nightly binned NEID RVs that were used in the analysis.

\subsubsection{The Habitable Zone Planet Finder at McDonald Observatory}

We observed GJ 3929 with the Habitable Zone Planet Finder \citep[HPF; ][]{mahadevan12, mahadevan14}, a near-infrared, (\(8080-12780\)\ \AA), high precision RV spectrograph. HPF is located at the 10 m Hobby-Eberly Telescope (HET) in Texas. HET is a fixed-altitude telescope with a roving pupil design. Observations on the HET are queue-scheduled, with all observations executed by the HET resident astronomers \citep{shetrone07}. HPF is fiber-fed, with separate science, sky and simultaneous calibration fibers \citep{kanodia18a}, and has precise, milli-Kelvin-level thermal stability \citep{stefansson16}. 

We extracted precise RVs with HPF using the modified version of \texttt{SERVAL} \cite{zechmeister18} optimized for use for HPF data as described in detail in \cite{stefansson20_a}. The RV reduction followed similar steps as outlined in Section \ref{sec:neidobservations}.

We obtained 18 observations of GJ 3929 with HPF over the course of 6 observing nights. These data were taken between 2021 August 27 and 2022 March 11. We obtained 3 consecutive exposures on each observing night, resulting in a median unbinned RV error of 7.15 m s$^{-1}$. Data taken on BJD = 2459649 were excluded from our analysis due to poor weather conditions. Our dataset then consists of 5 nightly binned HPF RVs. Due to the small quantity of the HPF data, we considered fits that did not utilize HPF data. We found that model results did not differ meaningfully whether HPF data were utilized, or not, and we include them in our final model for completeness. HPF spectra were still used to derive stellar parameters, as outlined in Section \ref{sec:stellar}.

\subsubsection{CARMENES RVs}

Our RV modeling also utilizes CARMENES RVs published in \cite{kemmer22}. \cite{kemmer22} published 78 high-precision RVs as a part of their study of GJ 3929 using the CARMENES spectrograph \citep{quirrenbach14}. CARMENES is a dual-channel spectrograph with visible and near-infrared (NIR) arms ($\mathcal{R}_{VIS}$ = 94600; $\mathcal{R}_{NIR}$ = 80400). CARMENES is located at the Calar Alto Observatory in Almería, Spain. RVs of GJ 3929 were taken between 2020 July 30 and 2021 July 19. Each observation lasted 30 minutes, with a median observation SNR of 74. 5 RVs were discarded due to a missing drift correction in \cite{kemmer22}, and we do so as well. This results in a final dataset containing 73 RVs. These RVs were taken using the visible arm of CARMENES, and have a median uncertainty of 1.9 m s$^{-1}$.

\section{Stellar Parameters}\label{sec:stellar}

We followed steps outlined in \cite{stefansson20_a} and \cite{beard22} to estimate $T_{\mathrm{eff}}$, $\log g$, and [Fe/H] values of GJ 3929. The \texttt{HPF-SpecMatch} code is based on the \texttt{SpecMatch-Emp} algorithm from \cite{yee17}, and compares the high resolution HPF spectrum of the target star of interest  to a library of high SNR as-observed HPF spectra. This library consists of slowly-rotating reference stars with well characterized stellar parameters from \cite{yee17} and an expanded selection of stars from \cite{mann15} in the lower effective temperature range. Our analysis was run on 2022 March 3, and the library contained 166 stars during our run.

We shift the observed target spectrum to a library wavelength scale and rank all of the targets in the library using a $\chi^2$ goodness-of-fit metric. After this initial $\chi^2$ minimization step, we pick the five best matching reference spectra. We then construct a weighted spectrum using their linear combination to better match the target spectrum. A weight is assigned to each of the five spectra according to its goodness-of-fit. We then assign the target stellar parameter $T_{\mathrm{eff}}$, $\log g$, and [Fe/H] values as the weighted average of the five best stars using the best-fit weight coefficients. The final parameters are listed in Table \ref{tab:stellartable}. These parameters were derived from the HPF order spanning 8670\AA - 8750\AA, as this order is cleanest of telluric contamination. We artificially broadened the library spectra with a $v \sin i$ broadening kernel \citep{gray92} to match the rotational broadening of the target star. We determined GJ 3929 to have a $v \sin i$ broadening value of $<$ 2 \unit{km/s}.

We used \texttt{EXOFASTv2} \citep{eastman13} to model the spectral energy distributions (SED) of GJ 3929 and to derive model-dependent constraints on the stellar mass, radius, and age. \texttt{EXOFASTv2} utilizes the BT-NextGen stellar atmospheric models \citep{allard12} during SED fits.  Gaussian priors were used for the 2MASS (\(JHK\)), Johnson (\(B V\)), and \textit{Wide-field Infrared Survey Explorer} magnitudes (WISE; $W1$, $W2$, $W3$, and $W4$) \citep[][]{wright10}. Our spectroscopically-derived host star effective temperature, surface gravity, and metallicity, were used as priors during the SED fits as well, and the estimates from \cite{bailer-jones21} were used as priors for distance. We further include in our priors estimates of Galactic dust by \cite{green19} to estimate the visual extinction, though we emphasize that this is a conservative upper limit: GJ 3929 is fairly close to Earth, and is likely to be foreground to much of the dust utilized in this estimate. We convert this upper limit to a visual magnitude extinction using the \(R_{v}=3.1\) reddening law from \cite{fitzpatrick99}. Our final model results are consistent with those derived in \cite{kemmer22}, and are visible in Table \ref{tab:stellartable}.

\begin{deluxetable*}{lccc}
\tablecaption{Summary of stellar parameters for GJ 3929. \label{tab:stellartable}}
\tablehead{\colhead{~~~Parameter}&  \colhead{Description}&
\colhead{Value}&
\colhead{Reference}}
\startdata
\multicolumn{4}{l}{\hspace{-0.2cm} \textbf{Main identifiers:}}  \\
~~~TOI & \tess{} Object of Interest & 2013 & \tess{} mission \\
~~~TIC & \tess{} Input Catalogue  & 188589164 & TICv8 \\
~~~GJ & Gliese-Jahreiss Nearby Stars & 3929 & Gliese-Jahreiss \\
~~~2MASS & \(\cdots\) & J15581883+3524236 & 2MASS  \\
~~~Gaia DR3 & \(\cdots\) & 1372215976327300480 & Gaia DR3\\
\multicolumn{4}{l}{\hspace{-0.2cm} \textbf{Equatorial Coordinates, Proper Motion and Spectral Type:}} \\
~~~$\alpha_{\mathrm{J2000}}$ &  Right Ascension (RA; deg) & 239.57754339(4) & Gaia DR3\\
~~~$\delta_{\mathrm{J2000}}$ &  Declination (Dec; deg) & 35.40815826(2) & Gaia DR3\\
~~~$\mu_{\alpha}$ &  Proper motion (RA; \unit{mas/yr}) &  -143.28 $\pm$ 0.07 & TICv8 \\
~~~$\mu_{\delta}$ &  Proper motion (Dec; \unit{mas/yr}) & 318.22 $\pm$ 0.08 & TICv8 \\
~~~$d$ &  Distance (pc) & 15.8 $\pm$ 0.02 & Bailer-Jones \\
\multicolumn{4}{l}{\hspace{-0.2cm} \textbf{Optical and near-infrared magnitudes:}}  \\
~~~$B$ & Johnson B mag & 14.333 $\pm$ 0.008 & TICv8\\
~~~$V$ & Johnson V mag & 12.67 $\pm$ 0.02 & TICv8\\
~~~$g^{\prime}$ &  Sloan $g^{\prime}$ mag  & 15.161 $\pm$ 0.006 & TICv8\\
~~~$r^{\prime}$ &  Sloan $r^{\prime}$ mag  & 12.2405 $\pm$ 0.0009 & TICv8 \\
~~~$i^{\prime}$ &  Sloan $i^{\prime}$ mag  & 10.921 $\pm$ 0.001 & TICv8 \\
 ~~~$T$  & \tess{} magnitude & 10.270 $\pm$ 0.007  & TICv8 \\
~~~$J$ & $J$ mag & 8.69 $\pm$ 0.02 & TICv8\\
~~~$H$ & $H$ mag & 8.10 $\pm$ 0.02 & TICv8\\
~~~$K_s$ & $K_s$ mag & 7.87 $\pm$ 0.02 & TICv8\\
~~~$W1$ &  WISE1 mag & 7.68 $\pm$ 0.02 & WISE\\
~~~$W2$ &  WISE2 mag & 7.54 $\pm$ 0.02 & WISE\\
~~~$W3$ &  WISE3 mag & 7.42 $\pm$ 0.02 & WISE\\
~~~$W4$ &  WISE4 mag & 7.27 $\pm$ 0.08 & WISE\\
\multicolumn{4}{l}{\hspace{-0.2cm} \textbf{Spectroscopic Parameters$^a$:}}\\
~~~$T_{\mathrm{eff}}$ &  Effective temperature in \unit{K} & 3384 $\pm$ 88 & This work\\
~~~$\mathrm{[Fe/H]}$ &  Metallicity in dex & -0.02 $\pm$ 0.12 & This work\\
~~~$\log(g)$ & Surface gravity (cm s$^{-2}$) & 4.89 $\pm$ 0.05 & This work\\
\multicolumn{4}{l}{\hspace{-0.2cm} \textbf{Model-Dependent Stellar SED and Isochrone fit Parameters$^b$:}}\\
~~~$M_*$ &  Mass ($M_{\odot}$) & 0.313$^{+0.027}_{-0.022}$ & This work\\
~~~$R_*$ &  Radius ($R_{\odot}$) & 0.32 $\pm$ 0.01 & This work\\
~~~$L_*$ &  Luminosity ($L_{\odot}$) & 0.0109$^{+0.0005}_{-0.0004}$ & This work\\
~~~$\rho_*$ &  Density ($\unit{g/cm^{3}}$) & 13.3 $\pm$ 1.1 & This work\\
~~~Age & Age (Gyr) & 7.1$^{+4.1}_{-4.9}$ & This work\\
~~~$A_v$ & Visual extinction (mag) & 0.005 $\pm$ 0.003 & This work\\
~~~$d$ & Distance (pc) & 15.822 $\pm$ 0.006 & This work\\
\multicolumn{4}{l}{\hspace{-0.2cm} \textbf{Other Stellar Parameters:}}           \\
~~~$v \sin i_*$ &  Rotational velocity (\unit{km/s})  & $<$ 2 & This work\\
~~~$\Delta RV$ &  ``Absolute'' radial velocity (\unit{km/s}) & 10.265 $\pm$ 0.008 & This work \\
~~~$U, V, W$ &  Galactic velocities (\unit{km/s}) & -21.05 $\pm$ 0.04,10.85 $\pm$ 0.06,14.66 $\pm$ 0.08 & Kemmer\\
\enddata
\tablenotetext{}{References are: TICv8 \citep{stassun18}, 2MASS \citep{cutri03}, Gaia DR3 (Gaia Collaboration et al. 2022j, in prep), Bailer-Jones \citep{bailer-jones18}, WISE \citep{wright10}, Kemmer \citep{kemmer22}}
\tablenotetext{a}{Derived using the HPF spectral matching algorithm from \cite{stefansson20_a}}
\end{deluxetable*}

\section{Analysis}\label{sec:analysis}

Both photometry and RV data were essential for characterizing GJ 3929, as the system may have two or more planets, though we have only detected transits of planet b. First, in Section \ref{sec:transitanalysis}, we investigate the transiting planet using our photometric data. Next, we analyze the RV data of GJ 3929 in Section \ref{sec:rvanalysis}. Then, we search for additional transiting signals. Finally, in Section \ref{sec:jointanalysis}, we combine both datasets to reach our final conclusion.

\subsection{Transit Analysis}\label{sec:transitanalysis}

A 2.6 day transit signal was originally identified by the MIT SPOC pipeline on 2020 June 19, then designated TOI-2013.01. Subsequently, \cite{kemmer22} confirmed the planetary nature of the signal in early 2022. We combine the TESS data with our follow-up transits in addition to other publicly available photometric data (detailed in Section \ref{sec:photometry}) to further refine the measured parameters of the system.

\subsubsection{Modeling the Photometry}

We modeled GJ 3929's photometry using the \textsf{exoplanet} software package \citep{exoplanet:joss}. First, we downloaded the TESS PDCSAP flux using \textsf{lightkurve} \citep{lightkurve}. We then performed a standard quality-flag filter, removing datapoints designated as of poor quality by the SPOC pipeline, and we median normalized the TESS data. We then combined the TESS data with our normalized ARCTIC and LCOGT data for joint analysis.

Initial fits to ARCTIC and LCOGT data appeared to have a slight residual trend, and so in our adopted fit we detrended ARCTIC and LCOGT photometry before combining the datasets. We utilized the \textsf{NumPy} polyfit function to fit a line for purposes of detrending \citep{harris20}. This function performs a simple least squares minimization to estimate the linear trend. This detrending was performed before modeling the data, as we found that including a detrending term in the model did not meaningfully improve our results, while increasing the complexity of our model.

We found it best to partition the photometric data into four regions of interest: the TESS data (which consist of two consecutive sectors), two different nights of ARCTIC data, and a night of LCOGT data. Due to the possibility of systematic offsets between nights, and the distinct conditions during each night of ARCTIC observations, we choose to treat each ARCTIC night separately in our model. Furthermore, the LCOGT data were taken with four different filters. Consequently, we model each filter separately. For each instrument-filter combination, then, we adopt a unique mean and jitter term. The mean terms are additive offsets to account for potential systematic shifts between nights, and are simply subtracted from all data points when fitting. The jitter terms are meant to model additional white noise not properly accounted for in the errorbars of the dataset, and are added in quadrature with the errorbars. Our model thus consists of seven total mean terms, and seven jitter terms.

The physical transit model was generated using \texttt{exoplanet} functions and the \texttt{starry} lightcurve package \citep{exoplanet:luger18}, which models the period, transit time, stellar radius, stellar mass, eccentricity, radius, and impact parameter to produce a simulated lightcurve. We adopt quadratic limb darkening terms to account for the change in flux that occurs when a planet approaches the limb of a star \citep{exoplanet:kipping13}. The two ARCTIC transits were taken using the same Semrock filter, and so we expect their limb-darkening behavior to be the same. Thus, we adopt the same limb-darkening parameters for each ARCTIC transit. We adopt distinct limb-darkening terms for the LCOGT data taken with the SDSS g$^{\prime}$, i$^{\prime}$, r$^{\prime}$, and z$_{s}^{\prime}$ filters. We note that this results in six pairs of limb darkening terms, in contrast to seven separate jitter and mean terms, but is physically motivated.

\begin{figure}[]
\centering
\includegraphics[width=0.5\textwidth]{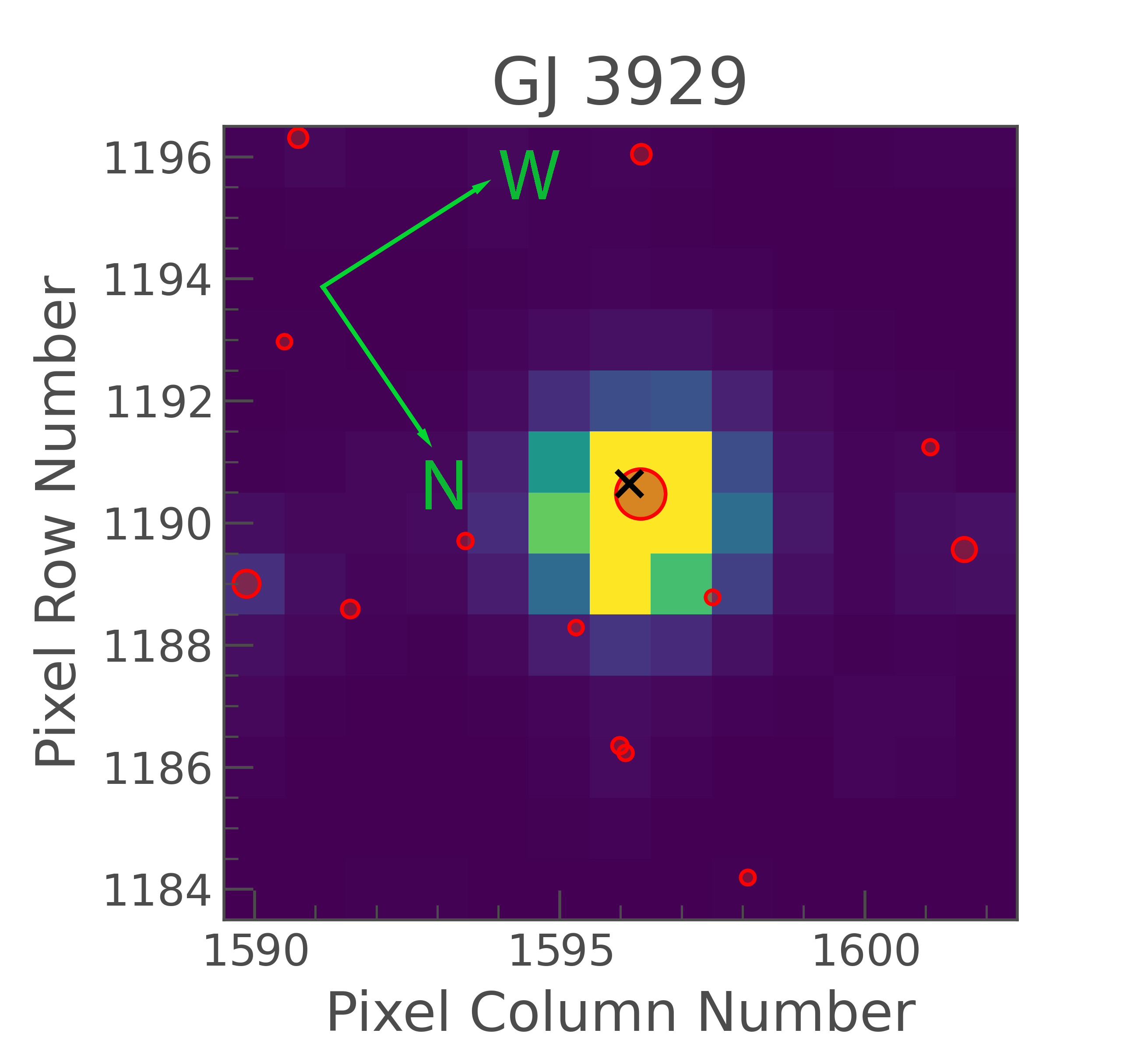}
\caption{TESS pixel image of GJ 3929 taken during Sector 24, created using the \texttt{eleanor} software package \citep{feinstein19}. The TICv8 position of GJ 3929 is indicated by a black x. Red circles correspond to Gaia resolved sources \citep{gaia_dr218}, with size corresponding to brightness. Because GJ 3929 does not have any bright neighbors, we do not use a dilution term.} \label{fig:tess_pixel}
\end{figure}

Similar to \cite{kemmer22}, we choose not to include a dilution term in our final model. GJ 3929 does not have many neighbors, and is much brighter than all of them (Figure \ref{fig:tess_pixel}). GJ 3929 has an estimated contamination ratio of 0.000765, meaning that 0.08$\%$ of its flux is possibly from nearby sources \citep{stassun19}. This suggests that a dilution term is not necessary.

\subsubsection{Inference}

After constructing a physical transit model using \texttt{starry}, we compare it to the data after it has been adjusted to account for offsets, and we add our jitter parameters in quadrature with the errorbars during likelihood estimation. Each free parameter is given a broad prior to prevent any biasing of the model, and we summarize the priors used in Table \ref{tab:priors}. The model is then optimized using \texttt{scipy.optimize.minimize} \citep{vertanen20}, which utilizes the Powell optimization algorithm \citep{powell98}. This optimization provides a starting guess for posterior inference. We then used a Markov Chain Monte Carlo (MCMC) sampler to explore the posterior space of each model parameter. \texttt{exoplanet} uses the Hamiltonian Monte Carlo (HMC) algorithm with a No U-Turn Sampler (NUTS) for increased sampling efficiency \citep{hoffman11}. We ran 10000 tuning steps and 10000 subsequent steps, and assessed convergence criteria using the Gelman-Rubin (G-R) statistic \citep{ford06}. We considered a chain well mixed if the G-R statistic was within 1$\%$ of unity. All the parameters in our model indicated convergence using this metric.

\begin{figure*}[] 
\centering
\includegraphics[width=\textwidth]{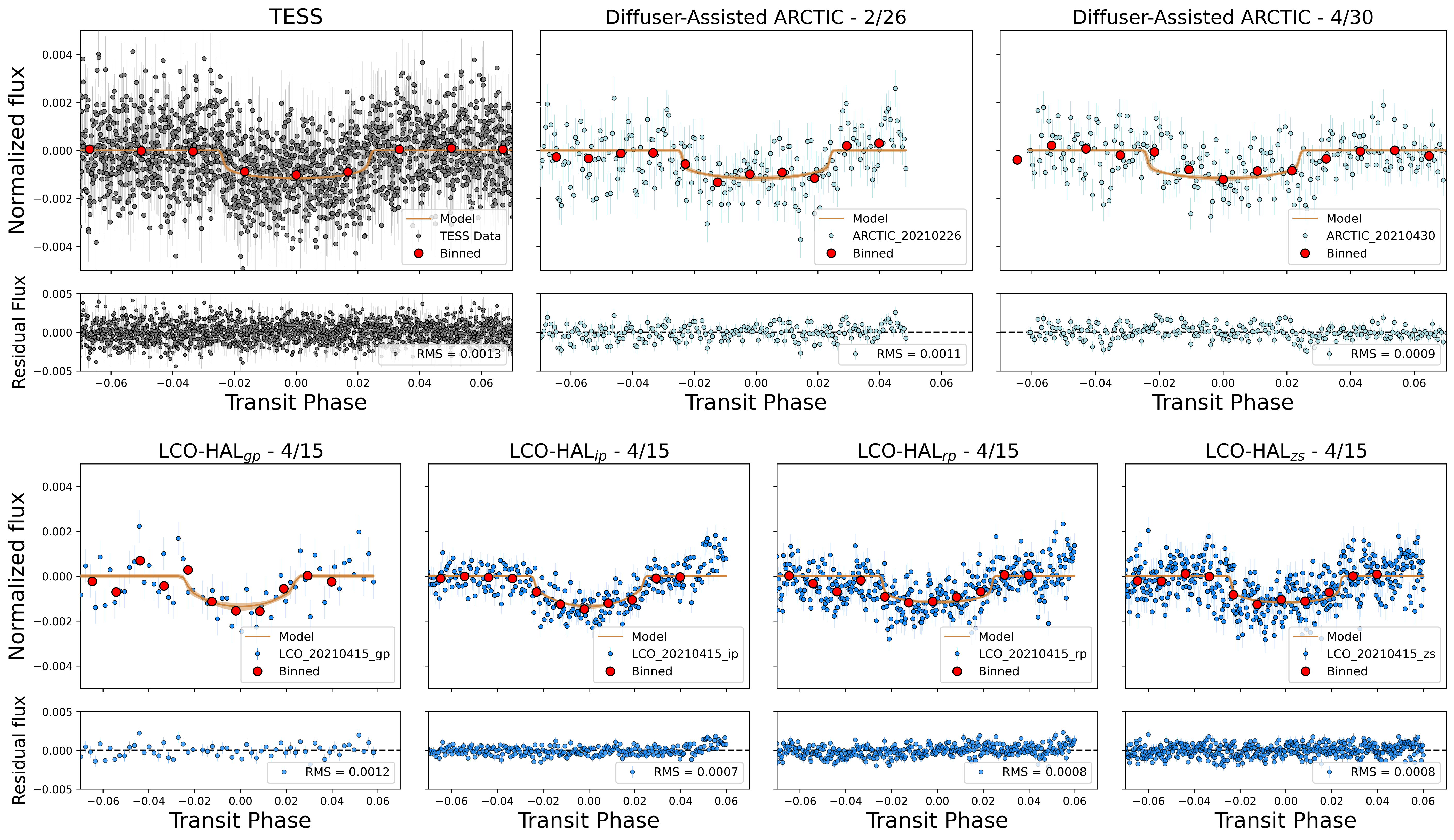}
\caption{Phase folded transit fits to TESS data, ARCTIC data, and LCOGT data. We separate the 2021 April 15 transit taken with LCOGT by filter and label them accordingly. Using all of these data allow us to modify previous radius estimates of GJ 3929b.} \label{fig:transit_final}
\end{figure*}

Our photometry-only fits are consistent with the joint fits adopted in Section \ref{sec:jointanalysis}. A final plot of the photometry, folded to the period of planet b is visible in Figure \ref{fig:transit_final}.

\subsection{Radial Velocity Analysis}\label{sec:rvanalysis}

\subsubsection{Periodogram Analysis}

\begin{figure*}[]
\centering
\includegraphics[width=\textwidth]{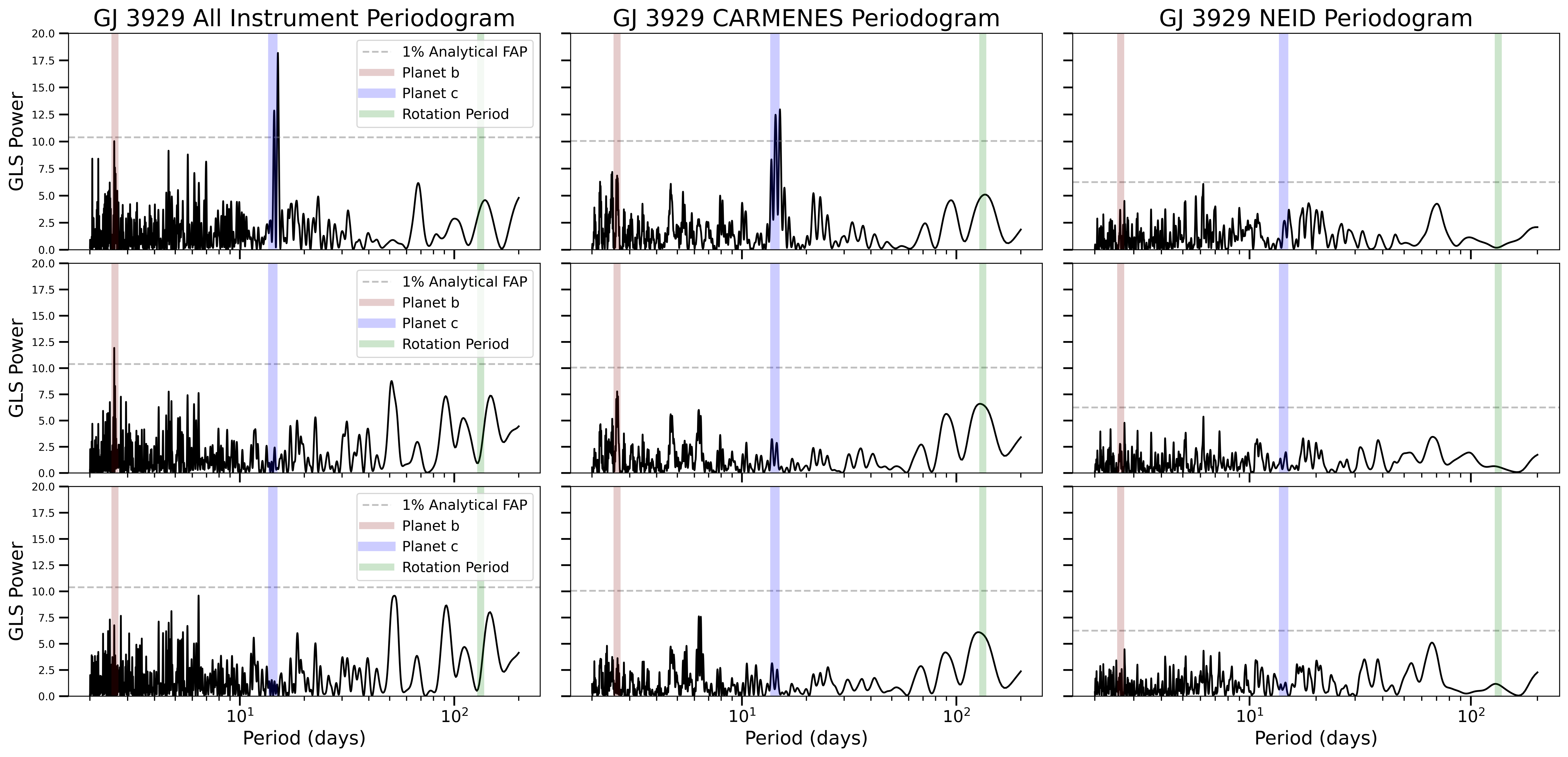}
\caption{Top: GLS periodograms of the combined dataset consisting of NEID, CARMENES, and HPF RVs, CARMENES data only, and NEID data only. Data have been adjusted for offsets. Middle: data after the subtraction of planet c, assuming the values derived in our final posterior fits. Bottom: GLS periodograms of data after the subtraction of planet b and planet c. We do not include a periodogram of HPF-only data due to its sparseness.} \label{fig:GLS}
\end{figure*}

We first used a Generalized Lomb-Scargle periodogram \citep[GLS;][]{zechmeister09} to analyze the RVs of GJ 3929, and to identify any periodic signals. We estimate the analytical false alarm levels and normalize the periodogram following the steps outlined in \cite{zechmeister09}, which assume Gaussian noise. With this assumption, we scale the sample variance (and false alarm levels) by $\frac{N - 1}{2}$ in order to reproduce the population variance, which is the quantity of interest in our analysis. Consistent with \cite{kemmer22}, we detected significant periodicities between 14-16 days. In contrast to \cite{kemmer22}, however, we find that when including the new, more precise NEID RVs (median CARMENES RV error $\sim$ 1.6 $\times$ median NEID RV error), as well as our HPF RVs, the 15 day signal has grown in power relative to the 14 day signal, suggesting that it might be the true signal. Relative peak strengths of alias frequencies in a periodogram do not always indicate the true period, however, and we detail a more formal model comparison later in the section. A plot of the combined-dataset periodogram, and periodograms on NEID and CARMENES only, are visible in Figure \ref{fig:GLS}. After the subtraction of the longer-period planet c, the signal of the 2.6 day planet b is clearly identifiable in the periodogram.

\subsubsection{Modeling the RVs}

We used the \texttt{RadVel} software package to analyze the RVs of GJ 3929 \citep{fulton18}. \texttt{RadVel} models an exoplanet's orbit by solving Kepler's equation using an iterative method outlined in \cite{danby88}. Each planetary orbit is then modeled by 5 fundamental parameters: the planet's orbital period ($P$), the planet's time of inferior conjunction ($T_{c}$), the eccentricity of the orbit ($e$), the argument of periastron ($\omega$), and the velocity semi-amplitude ($K$). We additionally include instrumental terms, $\gamma$ and $\sigma$, which account for systematic offsets between instruments, and excess white noise.

We construct the RV model in a Bayesian context, encoding prior information about each parameter as a part of the model. Similar to the fits described in Section \ref{sec:transitanalysis}, we adopt broad priors on the free parameters of our model to prevent any bias in our results. The primary exception being that during RV-only fits, we put tight priors on $P_{b}$ and $T_{con,b}$, as these are much more tightly constrained by transits than by RV fits. We emphasize, however, that our final adopted fit is a joint-fit between RVs and transits, detailed in Section \ref{sec:jointanalysis}. Detailed prior information is available in Table \ref{tab:priors}. 

\subsubsection{Inference}

In order to estimate the posterior probability of our model, we used an MCMC sampler to explore the posterior parameter space. \texttt{RadVel} utilizes the MCMC sampler outlined in \cite{foremanmackey13}. We first used the Powell optimization method to provide an initial starting guess for each parameter \citep{powell98}. We then ran 150 independent chains, and assessed convergence using the Gellman-Rubin statistic \citep[G-R;][]{ford06}. The sampling was terminated when the chains were sufficiently mixed. Chains are considered well-mixed when the G-R statistic for each parameter is $<$ 1.03, the minimum autocorrelation time factor is $\geq$ 75, the max relative change in autocorrelation time $\leq$ .01, and there are $\geq$ 1000 independent draws. All of our considered models eventually satisfied these conditions.

\begin{figure*}[h]
\centering
\includegraphics[width=\textwidth]{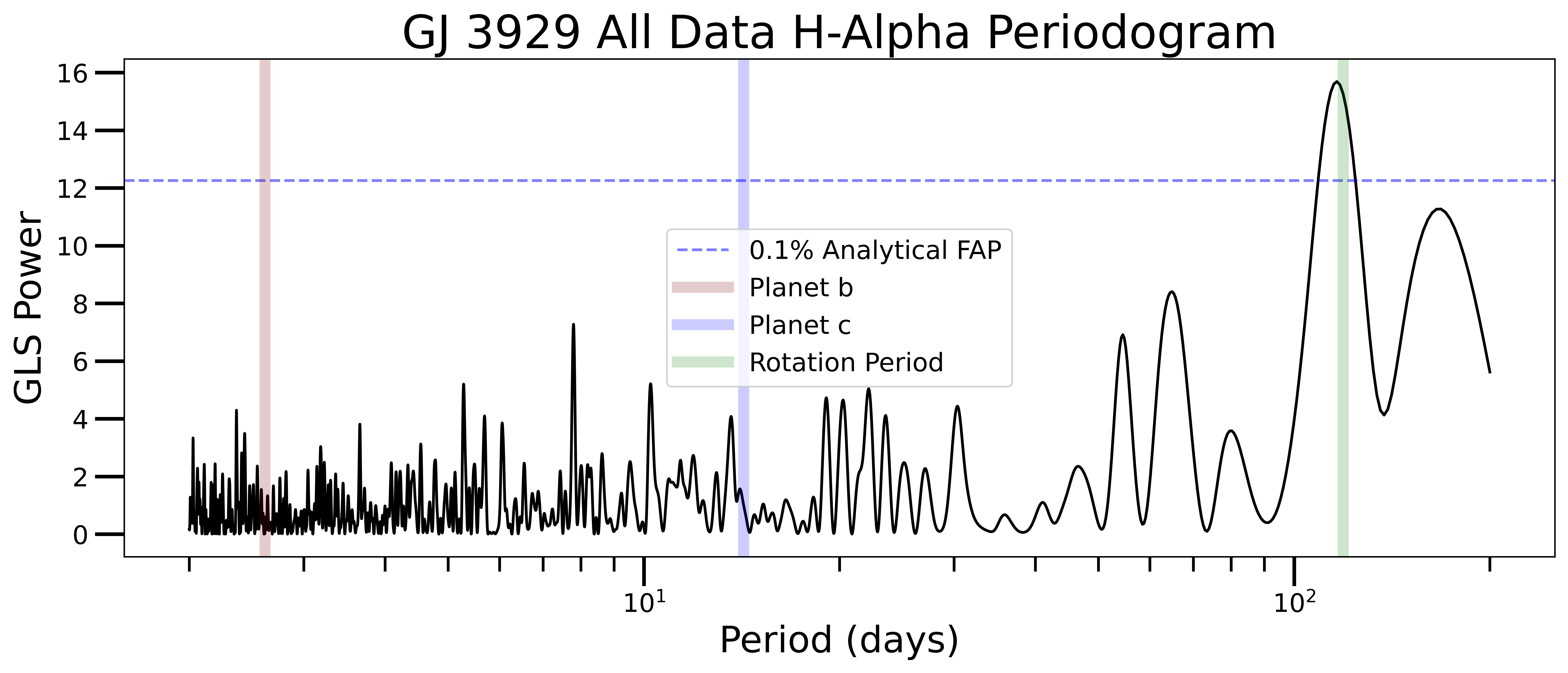}
\caption{GLS periodogram of the H$\alpha$ data taken by the CARMENES and NEID spectrographs. The only significant signal is at 116} days, which is most likely associated with the stellar rotation period identified in \cite{kemmer22}. Neither of the planetary periods has any significant power in this data. We do not include HPF, as its bandpass does not include the H$\alpha$ indicator. \label{fig:halpha}
\end{figure*}

We additionally considered the inclusion of a Gaussian Process \citep[GP;][]{ambikasaran15} model to account for coherent stellar activity. \cite{kemmer22} identify a rotation period of $\sim$ 120 days for GJ 3929. This value is derived from a combination of long-term photometry taken using the Hungarian Automated Telescope Network \citep[HATNet; ][]{bakos04}, the All-Sky Automated Search for SuperNovae \citep[ASAS-SN; ][]{shapee14}, and  Joan Oró Telescope \citep[TJO; ][]{colome10}, and periodogram analysis of the CARMENES H$\alpha$ values. We use the combined H$\alpha$ values from CARMENES and NEID to expand upon this, plotted in Figure \ref{fig:halpha}. While the maximum power occurs at a slightly shorter period than observed in \cite{kemmer22}, we note that rotational variability is often quasi-periodic in nature and periodograms can have trouble distinguishing longer periods \citep{lubin21}. Our value observed here is still consistent with the previously reported value, and we make no amendment to the system's rotation period.

The $>$ 100 day rotation period of this system is consistent with a quiet, slowly-rotating star, and we normally wouldn't expect a large RV signal due to activity. However, \cite{kemmer22} found an RV fit that included a GP to be preferred to an RV-only fit, and so we proceed with a series of fits, some of which include a GP. Our GP fits utilize the Quasi-Periodic GP kernel due to its flexibility and wide application in exoplanet astrophysics \cite[e.g.][]{haywood14,lopezmorales16}.

We also compared fits with the GP kernel that was adopted in \cite{kemmer22}. \cite{kemmer22} utilized a combination of two Simple Harmonic Oscillator \citep[SHO;][]{celerite2} kernels, outlined in more detail in \cite{kossakowski21}.

In order to explore the possibility of an additional planet in the GJ 3929 system, and the plausibility of stellar activity interfering with RV signals, we perform a model comparison. Model comparisons vary in the number of planets included, whether or not we include a GP to account for stellar noise, eccentric fits, and whether or not the second planet is modeled as the 14 day signal, or the 15 day signal. A full table of our model results is provided in Table \ref{tab:modelcomparison}. Our analysis found that both the Quasi-Periodic and dSHO GPs perform similarly in model comparison, and so we only include the Quasi-Periodic results for brevity. When comparing models, we use the Bayesian Information Criterion \citep[BIC;][]{kass95} and the Corrected Akaike Information Criterion \citep[AICc; ][]{hurvich87}. The BIC of each model can be used to estimate the Bayes Factor (BF), a measure of preference for one model over another. Half the difference in BIC between two models is used to estimate the Schwarz Criterion, which itself is an approximation of the $\log$ BF. The AICc is an approximation of the Kullback-Leibler information, another metric for ranking the quality of models \citep{hurvich87}.

\cite{kass95} suggest that a $\log_{10}$ BF $>$ 2 ($\ln$ BF $>$ 4.6) is decisive evidence for one model over another. For GJ 3929, our 2 planet ($\sim$15 day) model is preferred over the next best model, a 2 planet GP ($\sim$15 day), with a BF of 5.86 (\textsf{RadVel} estimates likelihoods using $\ln$), suggesting a strong preference for the no GP case. The AICc simply prefers the model that minimizes the AICc, which is also the 2 planet model ($\sim$15 day). Both methods of estimation are only asymptotically correct, but are preferred by a wide enough margin, and agree with one another. Consequently, we use these comparisons to justify selecting the 2 planet model ($\sim$15 day) without a GP as our best model.

\begin{deluxetable}{cccc}
\tablecaption{RV Model Comparisons$^{a}$ \label{tab:modelcomparison}}
\tabletypesize{\scriptsize}
\tablehead{\colhead{Fit}  &  \colhead{Number of Free}
& \colhead{BIC} & \colhead{AICc} \\
& \colhead{Parameters} & &}
\startdata
0 planet & 4 & 566.5757 & 552.0607 \\
----- & & \\
1 planet & 7 & 569.5241 & 548.4207 \\
1 planet ecc & 9 & 578.4362 & 553.4362 \\
1 planet GP & 11 & 574.0880 & 545.0023 \\
----- & & \\
2 planet ($\sim$14 day) & 10 & 560.2770 & 533.0948 \\
2 planet ($\sim$14 day) ecc (b) & 12 & 567.0928 & 536.1688 \\
2 planet ($\sim$14 day) ecc (c) & 12 & 562.6540 & 531.7300 \\
2 planet ($\sim$14 day) ecc (both) & 14 & 567.6228 & 533.2274 \\
2 planet GP ($\sim$14 day) & 14 & 576.62 & 542.2246 \\
\textbf{2 planet ($\sim$15 day)} & \textbf{10} & \textbf{545.5826} & \textbf{518.4004} \\
2 planet ($\sim$15 day) ecc (b) & 12 & 552.2589 & 521.3349 \\
2 planet ($\sim$15 day) ecc (c) & 12 & 551.8569 & 520.9229 \\
2 planet ($\sim$15 day) ecc (both) & 14 & 560.8289 & 525.8289 \\
2 planet GP ($\sim$15 day) & 14 & 557.3132 & 522.9178 \\
\enddata
\tablenotetext{a}{Model comparison was performed on RV-only fits. This is motivated in Section \ref{sec:jointanalysis}.}
\end{deluxetable}

\begin{figure*}[] 
\centering
\includegraphics[width=\textwidth]{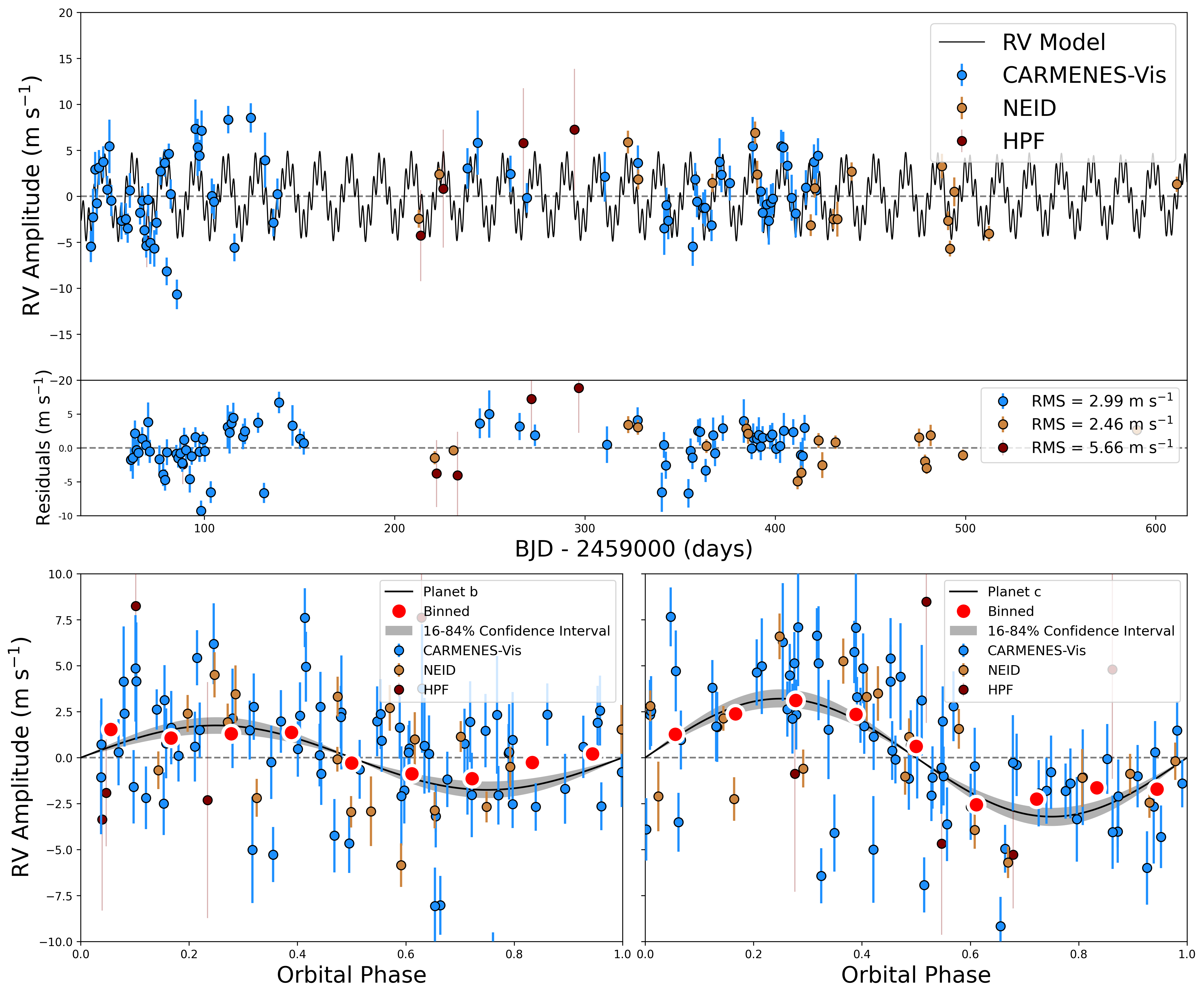}
\caption{Top: RV data of GJ 3929 used in our analysis. The data has been adjusted for systematic offsets. Overlaid in black is the 2 planet model. Bottom: Phase folds of our median fit to planets b and c after subtracting the other planet, with a 1$\sigma$ confidence interval overlaid. Jitter values are not included in the errors.} \label{fig:RV_final}
\end{figure*}

\subsection{An Additional Transiting Planet?}

\begin{figure*}[]
\centering
\includegraphics[width=\textwidth]{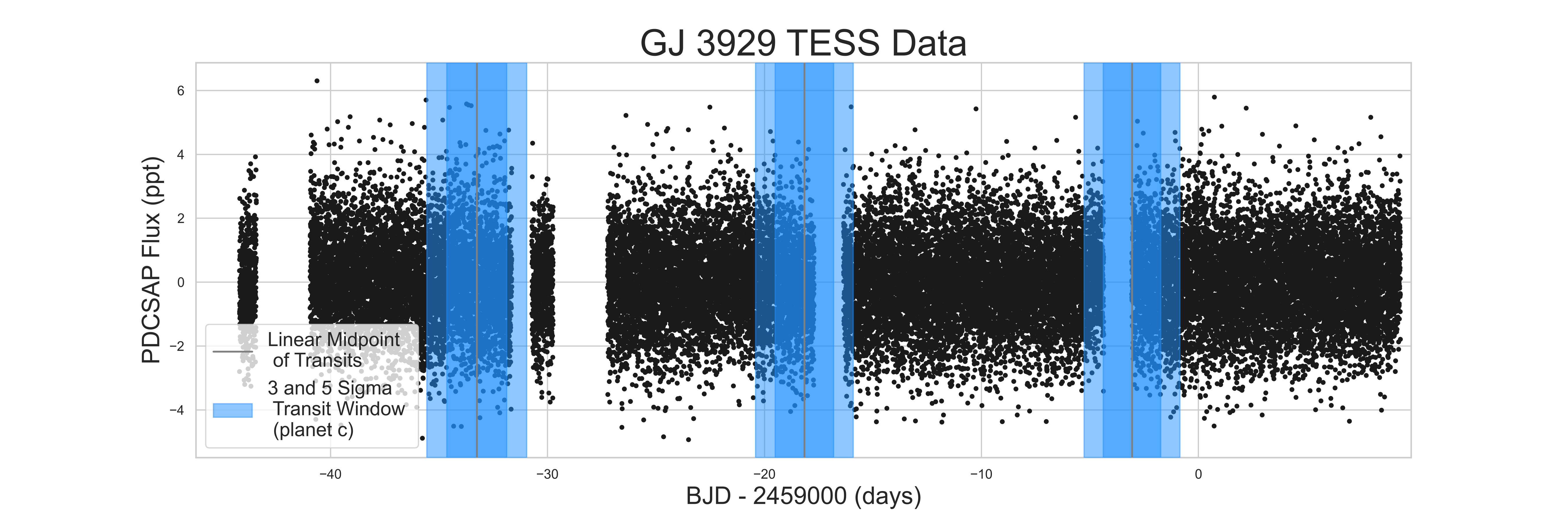}
\caption{TESS PDCSAP flux of GJ 3929, taken during Sectors 24 and 25. The projected linear ephemeris of planet c is marked by a vertical gray line, with the 3$\sigma$ and 5$\sigma$ windows of uncertainty overlaid. It seems plausible that the second and third transits of planet c might have fallen into data gaps, though the negative detection in the first transit window cannot easily be explained if planet c is transiting. Thus, we can rule out transits of planet c with 3$\sigma$ confidence.} \label{fig:transit_search_c}
\end{figure*}

As elaborated further in Section \ref{sec:rvanalysis} and detailed in \cite{kemmer22}, GJ 3929 RVs show two strong periodicities between 14-16 days. Consistent with \cite{kemmer22}, fitting either signal eliminates the other, suggesting that one is an alias of the other. Thus, we conclude that the two signals originate from a single source, though the true periodicity is originally unclear. Such a signal might be an additional planet, and if so, may be transiting. Here we search the TESS photometry for signs of additional transiting exoplanets, with a particular emphasis on planets in this period range.

We use the \texttt{TransitLeastSquares} \citep[TLS;][]{hippke19} python package in order to search for additional periodic transit signals in the TESS lightcurves. Unlike a Box-Least Squares algorithm \citep[BLS;][]{kovacs02}, which is used frequently in transit searches, the TLS adopts a more realistic transit shape, increasing its sensitivity to transiting exoplanets, especially smaller ones. Initially, we recover GJ 3929b with a signal detection efficiency (SDE) $>$ 35, a highly significant detection. \cite{dressing15} suggest that an SDE $>$ 6 represents a conservative cutoff for a ``significant" signal, though others adopt higher values \citep{siverd12,livingston18}. We then mask the transits of planet b, and continue the investigation. Our second check highlights a significant signal at 13.9 days (SDE = 12.74), somewhat close to the suspected planetary signals from the RVs. However, analysis of the candidate transit event itself seems inconsistent. Using the non-parametric mass-radius relationship from \texttt{mrexo} \citep{kanodia19}, we estimate that planet c would have a radius of 2.26 R$_{\oplus}$ using the minimum mass, and consequently a non-grazing transit depth of 4.19 ppt, more than 4 times as large as planet b. We caution that such mass-radius relationships are associated with a large uncertainty, though Figure \ref{fig:MR} makes it clear that GJ 3929c should at least be larger than planet b. However, this ``transit" observed by TLS at $\sim$ 14 days has a depth of 0.24 ppt. It is possible that the transits of this candidate are grazing, resulting in an anomalously small transit depth. However, the duration of the transits of this signal are also much longer than expected, at 0.42 days. This is not only inconsistent with a grazing transit, but would be too long for any transit at this period. Finally, the estimated transit phase is totally inconsistent with the time of conjunction found in \cite{kemmer22}. \textsf{TLS} finds a T$_{c}$ = 2459867 BJD, while \cite{kemmer22} would have expected a T$_{c}$ = 2459872 BJD (scaling back the time of conjunction reported). We thus conclude that this significant $\sim$ 14 day periodicity identified by the \texttt{TLS} package is not planet c, and is most likely noise.

It is possible that planet c is transiting, but that its transits fell into TESS data gaps. In Figure \ref{fig:transit_search_c}, we highlight where the transits of planet c would occur relative to the TESS photometry. We identify no clear transit signals in the data.

We calculate 3 and 5$
\sigma$ transit windows in Figure \ref{fig:transit_search_c} by using our posterior period and time of conjunction values for planet c, and back-propagating them using standard propagation of error. Consequently, from Figure \ref{fig:transit_search_c}, we can rule out non-grazing transits of planet c with 3$\sigma$ confidence.

\subsection{Candidate Planet, or Planet?}

\cite{kemmer22} designated the 14-15 day signal a planet candidate. While no transit signal is clearly detected at this period, we can rule out most false positive scenarios.

GJ 3929c might be a highly inclined binary or brown dwarf. While such a scenario cannot easily be ruled out, GJ 3929 has a Gaia RUWE value of 1.185, which is consistent with little astrometric motion \citep{gaia2016,gaia_dr3_21,lindegren21}. This suggests that a highly inclined binary scenario is unlikely.

Periodic or quasi-periodic RV signals can also be created by stellar magnetic activity. Our model comparison (Table \ref{tab:modelcomparison}) does not prefer a model that includes activity mitigation, and TESS photometry does not exhibit any obvious periodic variability (Figure \ref{fig:TESS}). Furthermore, no strong signal near 14 or 15 days exists in the H$\alpha$ indicator data (Figure \ref{fig:halpha}). The candidate rotation period does show up very strongly in the H$\alpha$ periodogram, however, and its value $>$ 100 days is far from either planet.

The $\sim$ 15 day signal associated with planet c is stable over the time baseline and across instruments, further suggesting a planetary explanation. Performing a 2 planet fit (without a GP) on the CARMENES data, and doing the same with all data, yields consistent results (K$_{c, carmenes} $ = 3.20 $\pm$ 0.58 m s$^{-1}$; K$_{c, all}$ = 3.18 $\pm$ 0.49 m s$^{-1}$). Planetary signals are expected to remain stable over any observational baseline, while activity-sourced signals increase or decrease in amplitude over time. This analysis provides additional evidence for the true period of planet c at $\sim$ 15 days. Performing the same analysis on the $\sim$ 14 day signal yields a noticeable decrease in amplitude with the new RV data (K$_{c, carmenes}$ = 2.64 $\pm$ 0.63 m s$^{-1}$; K$_{c, all}$ = 2.38 $\pm$ 0.52 m s$^{-1}$). While the two values are consistent, the 14 day signal appears more sensitive to the new data.

\subsection{Joint Transit-RV Analysis}\label{sec:jointanalysis}
\begin{deluxetable*}{llll}
\tablecaption{Priors Used for Bayesian Model Fits} \label{tab:priors}
\tabletypesize{\scriptsize}
\tablehead{\colhead{~~~Parameter Name} &
\colhead{Prior} &
\colhead{Units} & \colhead{Description}
}
\startdata
\sidehead{Planet b Orbital Parameters}
~~~P$_{b}$ & $\mathcal{U}^{a}(2.0, 3.0)$ & days & Period\\
~~~T$_{con,b}$ & $\mathcal{U}(2459319.0,2459320.0)$ & BJD (days) & Time of Inferior Conjunction \\ 
~~~$\sqrt{e}\cos\omega_{b}^{*}$ & $\mathcal{U}(-1,1)$ & ...  & Eccentricity Reparametrization \\
~~~$\sqrt{e}\sin\omega_{b}^{*}$ & $\mathcal{U}(-1,1)$ & ...  & Eccentricity Reparametrization \\
~~~$R_{p,b}/R_{*}$ & $\log\mathcal{N}^{b}(0.0953, 1.0)$ & ... & Scaled Radius \\
~~~b$_{b}$ & $\mathcal{U}(0.0, 1.0)$ & ...  & Impact Parameter \\
~~~K$_{b}$ & $\mathcal{U}(0.01,100)$ & m s$^{-1}$  & Velocity Semi-amplitude \\
\sidehead{Planet c Orbital Parameters}
~~~P$_{c}$ & $\mathcal{U}(14.5, 16)$ & days & Period\\
~~~T$_{con,c}$ & $\mathcal{U}(2459064.0, 2459080.0)$ & BJD (days) & Time of Inferior Conjunction \\ 
~~~$\sqrt{e}\cos\omega_{c}^{*}$ & $\mathcal{U}(-1,1)$ & ...  & Eccentricity Reparametrization \\
~~~$\sqrt{e}\sin\omega_{c}^{*}$ & $\mathcal{U}(-1,1)$ & ...  & Eccentricity Reparametrization \\
~~~K$_{c}$ & $\mathcal{U}(0.01,100)$ & m s$^{-1}$  & Velocity Semi-amplitude \\
\sidehead{GP Hyperparameters}
~~~$\eta_{1}^{*}$ & $\mathcal{U}(0,50)$ & m s$^{-1}$  & GP Amplitude \\
~~~$\eta_{2}^{*}$ & $\mathcal{U}(0.1,10000)$ & days  & Exponential Decay Length \\
~~~$\eta_{3}^{*}$ & $\mathcal{U}(100,150)$ & days  & Recurrence Rate (Rotation Period) \\
~~~$\eta_{4}^{*}$ & $\mathcal{U}(0.05,0.6)$ & ...  & Periodic Scale Length \\
\sidehead{Instrumental Parameters}
~~~$\gamma_{\rm{CARMENES}}$ & $\mathcal{U}(-100,100)$ & m s$^{-1}$ & CARMENES Systematic Offset \\
~~~$\gamma_{\rm{NEID}}$ & $\mathcal{U}(-100,100)$ & m s$^{-1}$ & NEID Systematic Offset \\
~~~$\gamma_{\rm{HPF}}$ & $\mathcal{U}(-100,100)$ & m s$^{-1}$ & HPF Systematic Offset \\
~~~$\sigma_{\rm{CARMENES}}$ & $\mathcal{U}(0.01,100)$ & m s$^{-1}$  & CARMENES Jitter \\
~~~$\sigma_{\rm{NEID}}$ & $\mathcal{U}(0.01,100)$ & m s$^{-1}$  & NEID Jitter \\
~~~$\sigma_{\rm{HPF}}$ & $\mathcal{U}(0.01,100)$ & m s$^{-1}$  & HPF Jitter \\
~~~$\sigma_{\rm{TESS}}$ & $\log\mathcal{N}(-9.48,2)$& ... & Photometric Jitter \\
~~~$\sigma_{\rm{ARCTIC-20210226}}$ & $\log\mathcal{N}(-9.67,2)$& ... & Photometric Jitter \\
~~~$\sigma_{\rm{ARCTIC-20210430}}$ & $\log\mathcal{N}(-11.88,2)$& ... & Photometric Jitter \\
~~~$\sigma_{\rm{LCO-HAL_{gp}}}$ & $\log\mathcal{N}(-12.41,2)$& ... & Photometric Jitter \\
~~~$\sigma_{\rm{LCO-HAL_{ip}}}$ & $\log\mathcal{N}(-13.17,2)$& ... & Photometric Jitter \\
~~~$\sigma_{\rm{LCO-HAL_{rp}}}$ & $\log\mathcal{N}(-12.96,2)$& ... & Photometric Jitter \\
~~~$\sigma_{\rm{LCO-HAL_{zs}}}$ & $\log\mathcal{N}(-12.53,2)$& ... & Photometric Jitter \\
~~~$\gamma_{\rm{TESS}}$ & $\mathcal{N}(0.0,10.0)$& ... & Photometric Offset \\
~~~$\gamma_{\rm{ARCTIC-20210226}}$ & $\mathcal{N}(0.0,10.0)$& ... & Photometric Offset \\
~~~$\gamma_{\rm{ARCTIC-20210430}}$ & $\mathcal{N}(0.0,10.0)$& ... & Photometric Offset \\
~~~$\gamma_{\rm{LCO-HAL_{gp}}}$ & $\mathcal{N}(0.0,10.0)$& ... & Photometric Offset \\
~~~$\gamma_{\rm{LCO-HAL_{ip}}}$ & $\mathcal{N}(0.0,10.0)$& ... & Photometric Offset \\
~~~$\gamma_{\rm{LCO-HAL_{rp}}}$ & $\mathcal{N}(0.0,10.0)$& ... & Photometric Offset \\
~~~$\gamma_{\rm{LCO-HAL_{zs}}}$ & $\mathcal{N}(0.0,10.0)$& ... & Photometric Offset \\
~~~$u_{\rm{TESS}}$ & $\mathcal{K}^{c} $ & ... & Quadratic Limb Darkening \\
~~~$u_{\rm{ARCTIC}}$ & $\mathcal{K}$& ... & Quadratic Limb Darkening \\
~~~$u_{\rm{LCO_{gp}}}$ & $\mathcal{K} $ & ... & Quadratic Limb Darkening \\
~~~$u_{\rm{LCO_{ip}}}$ & $\mathcal{K} $ & ... & Quadratic Limb Darkening \\
~~~$u_{\rm{LCO_{rp}}}$ & $\mathcal{K} $ & ... & Quadratic Limb Darkening \\
~~~$u_{\rm{LCO_{zs}}}$ & $\mathcal{K} $ & ... & Quadratic Limb Darkening \\
\enddata
\tablenotetext{a}{$\mathcal{U}$ is a uniform prior with $\mathcal{U}$(lower,upper)}
\tablenotetext{b}{$\mathcal{N}$ is a normal prior with $\mathcal{N}$(mean,standard deviation)}
\tablenotetext{c}{$\mathcal{K}$ is a reparametrization of a uniform prior for limb darkening, outlined in \citep{exoplanet:kipping13}}
\tablenotetext{*}{These parameters are not utilized in our final adopted fit. We include them for completeness.}
\normalsize
\end{deluxetable*}

The final step of our analysis is the combination of the transit fits and RV fits into one complete, joint analysis. We adopt this model as our best, final model as it is the most complete description of GJ 3929: it utilizes all data, and characterizes both planets that are observed in this system, while also characterizing properties of planet b that can only be gleaned from photometry, especially its radius. 

We performed a model comparison in Section \ref{sec:rvanalysis}, and we use that model comparison to select our preferred model, which is a 2 planet model without the use of a GP. We performed this model comparison in the RV analysis rather than the joint analysis for one primary reason: all the free parameters of interest are primarily measured in the RVs. First, we were interested in deciding between a 1 and 2 planet model. The second planetary signal is \textit{only} detected in the RVs; transit searches have been unsuccessful. Second, we wanted to differentiate between a 14 or 15 day period for planet c. Again, this signal is only represented in the RV data. Thirdly, we wanted to justify the use of a GP. Our primary consideration for the use of a GP was in the RVs, as the photometry are quiet, as expected. A $>$ 100 day rotation period would be unlikely to be observed in TESS PDCSAP flux, and the ground-based photometry are all far too short in baseline to be affected by a periodicity on even 1/100th of rotation period's timescale. Finally, we were interested in testing the veracity of eccentric models. Eccentricity, however, is much more strongly constrained by RVs than by photometry.

Our final, joint fit, then, was performed considering a 2 planet model, where the second planet period is constrained between 15 and 16 days in order to prevent the MCMC chains from clustering around the alias at 14.2 days. The model is circular, and we do not adopt any GP to account for excess noise. We use the \texttt{exoplanet} software package in the joint fit, and the transits are modeled identically as described in Section \ref{sec:transitanalysis}. The RVs are modeled in \texttt{exoplanet} as well, with two Keplerian orbital solutions that model both photometric and RV datasets simultaneously. In particular, the period and time of conjunction of each planet are shared between the datasets, while other orbital parameters are typically constrained to one dataset or another. A full list of the priors used in our model are available in Table \ref{tab:priors}.

We again use the HMC algorithm with a NUTS sampler for increased sampling efficiency. We again run 10000 tuning steps and 10000 subsequent steps posterior estimation steps. Our final transit fits are visible in Figure \ref{fig:transit_final}. Our final RV fit is visible in Figure \ref{fig:RV_final}. Finally, our posterior estimates for each model free parameter are listed in Table \ref{tab:posteriors}.

\begin{deluxetable*}{llcc}
\tablecaption{Derived Parameters for both planets \label{tab:posteriors}}
\tablehead{\colhead{~~~Parameter} &
\colhead{Units} &
\colhead{GJ 3929b} & \colhead{GJ 3929c}
}
\startdata
\sidehead{\textbf{Orbital Parameters:}}
~~~Orbital Period $\dotfill$ & $P$ (days) $\dotfill$ & 2.616235 $\pm$ 0.000005 & 15.04 $\pm$ 0.03 \\
~~~Time of Inferior Conjunction $\dotfill$ & $T_C$ (BJD$\textsubscript{TDB}$) $\dotfill$ & 2458956.3962 $\pm$ 0.0005 & 2459070.9 $\pm$ 0.4 \\
~~~Eccentricity $\dotfill$ & $e$ $\dotfill$ & 0 (fixed) & 0 (fixed)\\
~~~Argument of Periastron $\dotfill$ & $\omega$ (degrees) $\dotfill$ & 90 (fixed) & 90 (fixed) \\
~~~RV Semi-Amplitude $\dotfill$ & $K$ (m/s) $\dotfill$ &
 1.77$^{+0.44}_{-0.45}$ & 3.22 $\pm$ 0.51 \\
\sidehead{\textbf{Transit Parameters:}}
~~~Scaled Radius $\dotfill$ & $R_{p}/R_{*}$ $\dotfill$ & 
0.0156 $\pm$ 0.0003 & ... \\
~~~Scaled Semi-major Axis $\dotfill$ & $a/R_{*}$ $\dotfill$ & 16.8 $\pm$ 0.5 & ... \\
~~~Impact Parameter $\dotfill$ & $b$ $\dotfill$ & 0.11$^{+0.06}_{-0.07}$ & ... \\
~~~Orbital Inclination $\dotfill$ & $i$ (degrees) $\dotfill$ & 88.442 $\pm$ 0.008 & ... \\
~~~Transit Duration $\dotfill$ & $T_{14}$ (days) $\dotfill$ &  0.0495$^{+0.0008}_{-0.0007}$ & ... \\
\sidehead{\textbf{Planetary Parameters:}}
~~~Mass $\dotfill$ & $M_{p}$ (M$_\oplus$) $\dotfill$ & 1.75$^{+0.44}_{-0.45}$ & 5.71$^{a}\pm$ 0.94 \\
~~~Radius$\dotfill$ & $R_{p}$  (R$_\oplus$) $\dotfill$& 1.09 $\pm$ 0.04 & ... \\
~~~Density $\dotfill$ & $\rho_{p}$ (g/$\unit{cm^{3}}$)$\dotfill$ & 7.3 $\pm$ 2.0
 & ... \\
~~~Semi-major Axis$\dotfill$ & $a$ (AU) $\dotfill$ & 0.0252 $\pm$ 0.0005 & 0.081 $\pm$ 0.002 \\
~~~Average Incident Flux$\dotfill$ & $\langle F \rangle$ ($\unit{W/m^2}$)$\dotfill$ & 24000 $\pm$ 1000 & 2300 $\pm$ 100 \\
~~~Planetary Insolation$\dotfill$ & $S$ (S$_\oplus$)$\dotfill$ & 17.3 $\pm$ 0.7 & 1.68 $\pm$ 0.07 \\
~~~Equilibrium Temperature$^{b}$ $\dotfill$ & $T_{\mathrm{eq}}$ (K)$\dotfill$ & 568 $\pm$ 6 & 317 $\pm$ 3 \\
\sidehead{\textbf{Instrumental Parameters}}
~~~RV Jitter $\dotfill$ & $\sigma_{CARMENES}$ (m/s) $\dotfill$ & 
1.80 $\pm$ 0.48 &  \\
 & $\sigma_{NEID}$ (m/s) $\dotfill$ & 
2.25 $\pm$ 0.66 &  \\
 & $\sigma_{HPF}$ (m/s) $\dotfill$ & 
6 $\pm$ 7 &  \\
~~~RV Offset $\dotfill$ & $\gamma_{CARMENES}$ (m/s) $\dotfill$ & 
0.97 $\pm$ 0.39 &  \\
 & $\gamma_{NEID}$ (m/s) $\dotfill$ & 
5.56 $\pm$ 0.66 &  \\
 & $\gamma_{HPF}$ (m/s) $\dotfill$ & 
8 $\pm$ 4 &  \\
~~~Limb Darkening $\dotfill$ & $u_{1,\rm{TESS}}$, $u_{2,\rm{TESS}}$ $\dotfill$ & 0.3$^{+0.3}_{-0.2}$,0.3 $\pm$ 0.4 &  \\
 & $u_{1,\rm{ARCTIC}}$, $u_{2,\rm{ARCTIC}}$ $\dotfill$  & 0.5 $\pm$ 0.3,0.0$^{+0.4}_{-0.3}$ &  \\
  & $u_{1,\rm{LCO-HALgp}}$, $u_{2,\rm{LCO-HALgp}}$ $\dotfill$ & 1.0$^{+0.5}_{-0.6}$,-0.3$^{+0.5}_{-0.4}$ &  \\
  & $u_{1,\rm{LCO-HALip}}$, $u_{2,\rm{LCO-HALip}}$ $\dotfill$ & 1.0$^{+0.3}_{-0.4}$,-0.3$^{+0.4}_{-0.3}$ &  \\
  & $u_{1,\rm{LCO-HALrp}}$, $u_{2,\rm{LCO-HALrp}}$ $\dotfill$ & 0.3$^{+0.3}_{-0.2}$,0.3 $\pm$ 0.4 &  \\
  & $u_{1,\rm{LCO-HALzs}}$, $u_{2,\rm{LCO-HALzs}}$ $\dotfill$ & 0.3$^{+0.3}_{-0.2}$,0.3 $\pm$ 0.4 &  \\
~~~Photometric Jitter $\dotfill$ & $\sigma_{\rm{TESS}}$ (ppm) $\dotfill$ & 10$^{+14}_{-7}$ &  \\
 & $\sigma_{\rm{ARCTIC-20210226}}$ (ppm) $\dotfill$  & 514 $\pm$ 100 &  \\
 & $\sigma_{\rm{ARCTIC-20210430}}$ (ppm) $\dotfill$  & 545 $\pm$ 60 &  \\
 & $\sigma_{\rm{LCO-HALgp}}$ (ppm) $\dotfill$ & 4$^{+33}_{-4}$ &  \\
 & $\sigma_{\rm{LCO-HALip}}$ (ppm) $\dotfill$ & 356 $\pm$ 42 &  \\
 & $\sigma_{\rm{LCO-HALrp}}$ (ppm) $\dotfill$ & 558 $\pm$ 40 &  \\
 & $\sigma_{\rm{LCO-HALzs}}$ (ppm) $\dotfill$ & 480 $\pm$ 40 &  \\
~~~Photometric Mean $\dotfill$ & mean$_{\rm{TESS}}$ (ppm) $\dotfill$ & 40 $\pm$ 8 &  \\
 & mean$_{\rm{ARCTIC-20210226}}$ (ppm)$\dotfill$  & 400 $\pm$ 70 &  \\
 & mean$_{\rm{ARCTIC-20210430}}$ (ppm)$\dotfill$  & 340 $\pm$ 60 &  \\
 & mean$_{\rm{LCO-HALgp}}$ (ppm) $\dotfill$ & 350 $\pm$ 100 & \\
 & mean$_{\rm{LCO-HALip}}$ (ppm) $\dotfill$ & 360 $\pm$ 40 & \\
 & mean$_{\rm{LCO-HALrp}}$ (ppm) $\dotfill$ & 350 $\pm$ 40 & \\
 & mean$_{\rm{LCO-HALzs}}$ (ppm) $\dotfill$ & 340 $\pm$ 40 & \\
\enddata
\tablenotetext{a}{Minimum mass}
\tablenotetext{b}{Estimated assuming an albedo of 0}

\normalsize
\end{deluxetable*}

\section{Discussion}\label{sec:discussion}

We have refined the measured parameters for GJ 3929b (P$_{b}$ = 2.616235 $\pm$ 0.000005 days; R$_{b}$ = 1.09 $\pm$ 0.04 R$_{\oplus}$; M$_{b}$ = 1.75$^{+0.44}_{-0.45}$ M$_{\oplus}$; $\rho_{b}$ = 7.3 $\pm$ 2.0 g cm$^{-3}$) and GJ 3929c (P$_{c}$ = 15.04 $\pm$ 0.03 days; M$\sin i_{c}$ = 5.71 $\pm$ 0.94 M$_{\oplus}$). 

GJ 3929 joins a growing list of M dwarf systems that contain a short period terrestrial planet, accompanied by a non-transiting, more massive planet \citep[i.e.][]{bonfils18}. Additionally, the possible existence of additional planetary companions cannot be ignored. M dwarfs in particular tend to have higher multiplicity of smaller exoplanets. \cite{lu20} used metallicity correlations when studying M dwarf systems to estimate how much planet-forming material is present in an initial planetary disk. It is likely that a correlation exists between metallicity of the host star and the amount of planet-forming material in a disk, especially for late-type stars \citep{bonfils05,johnson09}. \cite{lu20} estimate only 9 M$_{\oplus}$ of material for forming planets in a metal-poor ([Fe/H] = -0.5) early M dwarf (M$_{*}$ = 0.6 M$_{\odot}$). While GJ 3929 is smaller (M$_{*}$ = 0.32 M$_{\odot}$) than this system, its metallicity is much closer to the Sun ([Fe/H] = -0.05), giving it $\sim$ 15 M$_{\oplus}$ of material to form planets, if we assume the disk-to-star mass ratio of 0.01 that \cite{lu20} adopt. The sum of the median mass of GJ 3929b (1.75 M$_{\oplus}$) and the median minimum mass of GJ 3929c (5.70 M$_{\oplus}$) is significantly less than this value, implying that either planet c is significantly inclined and much more massive than we estimate, additional planets exist in the system, or the extra disk material was accreted onto the star. 

\begin{figure}[] 
\centering
\includegraphics[width=0.48\textwidth]{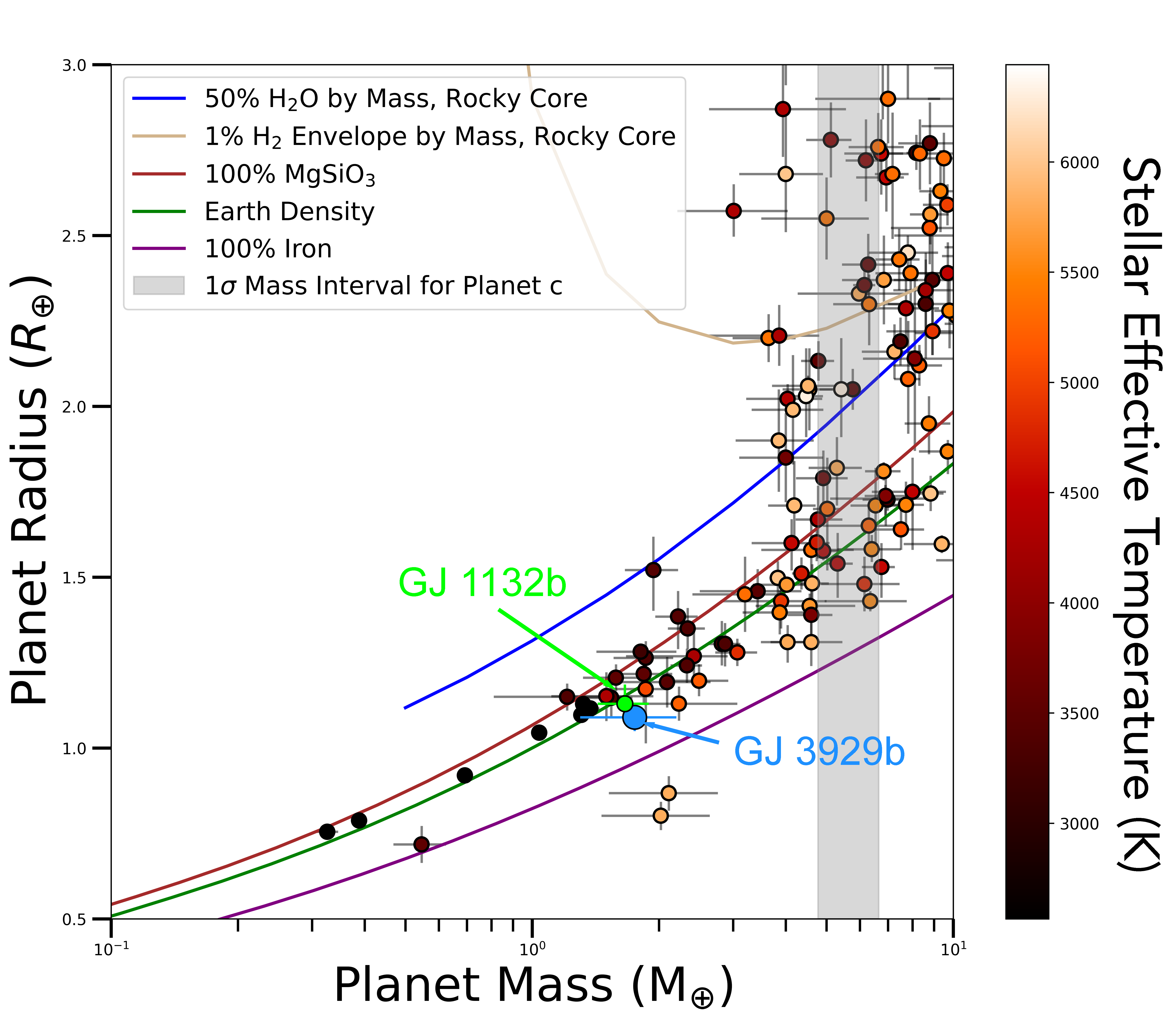}
\caption{Mass-Radius diagram of exoplanets taken from the NASA Exoplanet Archive on 2022 April 5. We restricted our study to planets with measured radii and masses. Colors indicate the stellar effective temperature of the system's host star. GJ 3929b is depicted in blue, and a region spanning the possible positions of planet c is visible in gray. We also include GJ 1132b in green, as it is a similar system discussed further in the text. A few theoretical density estimates are included as outlined in \cite{zeng19}.} \label{fig:MR}
\end{figure}

We highlight the GJ 1132 system, characterized in \cite{bonfils18}, for comparison with GJ 3929. GJ 1132b is also a short period, Earth-sized rocky planet orbiting an M dwarf, with an additional non-transiting companion. GJ 3929b is denser than GJ 1132b, as seen in Figure \ref{fig:MR}, though it's longer orbital period makes its RV semi-amplitude a bit smaller. We include comparisons to this system further in the discussion to help frame GJ 3929 in the context of similar systems.

\subsection{Planet b}

GJ 3929b is an Earth-sized exoplanet, placing it below the radius gap for M Dwarfs \citep{vaneylen21, petigura22}. Our mass and radius estimates allow us to constrain GJ 3929b's bulk density, and confirm its consistency with a composition slightly denser than Earth (Figure \ref{fig:MR}). Due to its proximity to its host star, GJ 3929b probably lost much of its atmosphere due to XUV flux \citep{vaneylen18}. The addition of a non-transiting second planet in the system originally confounded our RV analysis of the system, and further emphasizes the challenges discussed in \cite{he21} relating to the mass measurement of transiting planets. Since more than half of the time, the transiting planet in a system with non-transiting companions does not have the largest semi-amplitude, initial follow-up can be confusing.

GJ 3929b is Venus-like (S$_{b}$ = 17.3$^{+0.8}_{-0.7}$), in that it resides in its host-star's Venus-zone. This is defined as the boundary between the runaway-greenhouse inner edge of the Habitable Zone \citep{kasting93, kopparapu13, kopparapu14}, and an orbital distance that would produce 25 times Earth-like flux \citep{kane14,ostberg19}. Learning more about planets in the Venus-zone is an important step towards discovering Earth-twins. Spectroscopic observations of the Solar System, for example, would have a hard time distinguishing between Earth and Venus, despite their drastically different surface environments \citep{jordan21}. GJ 3929b is an excellent planet for studying the differences in spectra for a system that is Venus-like, and for which we are certain that it is nothing like Earth.

Fortunately, GJ 3929b is amenable to atmospheric study with the James Webb Space Telescope \citep[JWST; ][]{gardner06}. Beyond learning more about exo-Venuses, studying the atmosphere of GJ 3929b could help reveal the evolutionary history of the system, and shed light on planet formation models. GJ 3929b has an estimated Transmission Spectroscopy Metric \citep[TSM; ][]{kempton18} of 14 $\pm$ 4, placing it in the top quintile of Earth-sized exoplanets amenable to JWST observations. The density of GJ 3929b does not suggest a thick atmosphere, though a thin atmosphere of outgassed volatiles, a thin atmosphere lacking in volatiles and consisting of silicates and enriched in refractory elements, or a no-atmosphere scenario are all plausible \citep{seager10}.

\begin{figure}[] 
\centering
\includegraphics[width=0.48\textwidth]{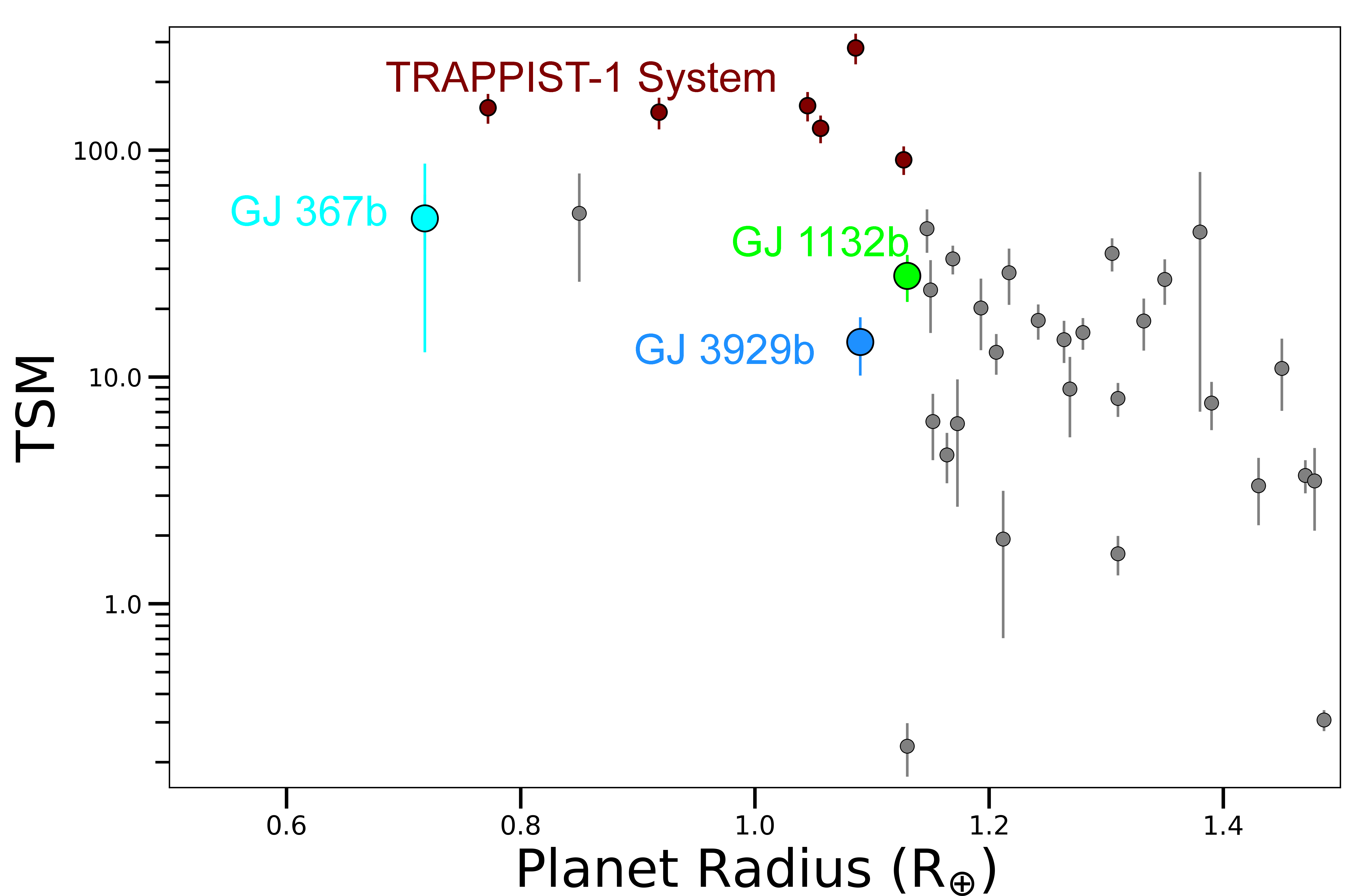}
\caption{Transmission Spectroscopy Metric \cite[TSM;][]{kempton18} of various planets taken from the NASA Exoplanet Archive on 2022 April 5. We note that GJ 3929b is in a sparsely populated region of parameter space, due largely to the difficulty of studying small exoplanets. We highlight a few other small-planet systems that are amenable to transmission spectroscopy.} \label{fig:TSM}
\end{figure}

In Figure \ref{fig:TSM}, we highlight GJ 3929b's TSM in the context of other small exoplanets. We include all exoplanets with sufficient information to calculate a TSM on the NASA Exoplanet Archive, though we caution that only exoplanets with $>$ 3$\sigma$ mass measurements are likely to see follow-up with JWST due to a degeneracy in the interpretation of spectra \citep{batalha19}. GJ 3929b occupies a truly rare position in this space, as quality mass measurements are very challenging for planets of its size, and small planets with mass measurements are usually not very amenable to transmission spectroscopy. We highlight a few other small planets amenable to transmission spectroscopy. Besides the TRAPPIST-1 system \citep{agol21}, which is exceptional in most parameter spaces, few small planets are better for transmission spectroscopy than GJ 3929b. While GJ 1132b is a similar system to GJ 3929b, and its TSM is slightly larger, GJ 3929b is brighter, making high-SNR measurements with JWST more likely, and making it more attractive for ground-based follow up. On the other hand, GJ 367b is an ultra-short period (USP) planet with a much higher TSM than GJ 3929b. However, its USP nature makes the existence of an atmosphere far less likely than for GJ 3929b, and further any such atmosphere would likely exhibit very different chemistries from GJ 3929b, since its equilibrium temperature is more than 3 times hotter  \citep[T$_{eq, GJ367b}$ = 1745 $\pm$ 43; K][]{lam21}.

\subsection{Planet c}

It is not clear whether or not GJ 3929c is a transiting exoplanet, though we detect no transits of this system in this study. Consequently, we cannot measure the radius of planet c, nor its bulk density.

The measured minimum mass of GJ 3929c suggests that it is at least a sub-Neptune in size when predicted from the mass-radius relationship, and perhaps larger \citep{kanodia19}. M dwarf systems consisting of a close-in, terrestrial exoplanet and longer period sub-Neptunes are common occurrences \citep{rosenthal21,sabotta21}, though the brightness of GJ 3929 allows for a more detailed study than is often the case. GJ 3929 will not be observed by TESS during Cycle 5, though the success of the TESS mission suggests that it will likely continue for years longer. Additionally, the advent of future photometric missions \citep[i.e.~PLATO; ][]{magrin18} suggest that GJ 3929 will probably receive additional photometric observations in the future, and a transit of planet c may someday be identified.

\subsection{Comparison to Kemmer et al. (2022)}

The addition of HPF and NEID RV data, as well as diffuser-assisted ARCTIC data have refined or changed various measured and derived parameters for each planet. Furthermore, our choice to use the $\sim$ 15 day signal as the period of GJ 3929c has an additional effect on several of the qualities of the planet.

The period and transit time of planet b are fully consistent with those found in \cite{kemmer22}, though the uncertainty is slightly larger in our case. This is most likely due to \cite{kemmer22}'s use of more transit data and in general modeling more transits of planet b. We prioritized higher precision photometry, and consequently opted not to use the SAINT-EX photometry or the additional LCO data utilized in \cite{kemmer22}. Furthermore our team did not have access to the transits obtained by the Observatorio de Sierra Nevada (OSN). Our additional ARCTIC photometry changed the radius measurement from 1.150 $\pm$ 0.04 R$_{\oplus}$ to 1.09 $\pm$ 0.04 R$_{\oplus}$, though we note that these values are 1$\sigma$ consistent.

The additional RVs did not shrink the formal 1$\sigma$ errorbars of the measured RV semi-amplitudes, but did modify the mean posterior values and the resulting K/$\sigma$ of our mass measurements are improved. For planet b, \cite{kemmer22} found a mass of 1.21$^{+0.40}_{-0.42}$ M$_{\oplus}$, and we find a mass of 1.75$^{+0.44}_{-0.45}$ M$_{\oplus}$. Similarly for planet c, \cite{kemmer22} found a minimum mass of 5.27$^{+0.74}_{-0.76}$ M$_{\oplus}$, while we measure a minimum mass of 5.71 $\pm$ 0.94 M$_{\oplus}$. We note, however, that changing the period of planet c likely played a role in this change as well, not merely the additional RVs.

Perhaps the most significant departure from \cite{kemmer22} is that our final model did not utilize a GP. In fact, this is probably the most significant contribution to the increased mass uncertainties in our fits. When utilizing a GP, our model does yield more precise mass uncertainties than those in \cite{kemmer22}, which is expected due to our inclusion of additional data. The increased amplitudes remain, however, suggesting that their difference is not related to the use of a GP. As shown in Table \ref{tab:modelcomparison}, we cannot justify the use of a GP in our final fit.

\section{Summary}\label{sec:summary}

We use RVs from the NEID, HPF, and CARMENES spectrographs to characterize the transiting planet GJ 3929b, and the probably non-transiting planet GJ 3929c. We use diffuser-assisted photometry from the ARCTIC telescope in combination with LCOGT and TESS photometry in order to improve the radius of GJ 3929b (R$_{b}$ = 1.09 $\pm$ 0.04 R$_{\oplus}$), and we use RVs from CARMENES, NEID, and HPF to measure the mass of both planets (M$_{b}$ = 1.75 $\pm$ 0.45 M$_{\oplus}$; M$\sin i_{c}$ = 5.70 $\pm$ 0.92 M$_{\oplus}$). We conclude that GJ 3929 is a 2 planet system with a 2.61626 $\pm$ 0.000005 day transiting exo-Venus that is highly amenable to transmission spectroscopy. GJ 3929c is a more massive planet orbiting with a period of 15.04 $\pm$ 0.03 days that is unlikely to transit. 

\section{Acknowledgements}

This paper includes data collected by the TESS mission. Funding for the TESS mission is provided by the NASA's Science Mission Directorate.

The Hobby-Eberly Telescope (HET) is a joint project of the University of Texas at Austin, the Pennsylvania State University, Ludwig-Maximilians-Universität München, and Georg-August-Universität Göttingen. The HET is named in honor of its principal benefactors, William P. Hobby and Robert E. Eberly.

The authors thank the HET Resident Astronomers for executing the observations included in this manuscript.

We would like to acknowledge that the HET is built on Indigenous land. Moreover, we would like to acknowledge and pay our respects to the Carrizo $\&$ Comecrudo, Coahuiltecan, Caddo, Tonkawa, Comanche, Lipan Apache, Alabama-Coushatta, Kickapoo, Tigua Pueblo, and all the American Indian and Indigenous Peoples and communities who have been or have become a part of these lands and territories in Texas, here on Turtle Island.

These results are based on observations obtained with the Habitable-zone Planet Finder Spectrograph on the HET. The HPF team was supported by NSF grants AST-1006676, AST-1126413, AST-1310885, AST-1517592, AST-1310875, AST-1910954, AST-1907622, AST-1909506, ATI 2009889, ATI-2009982, AST-2108512, AST-2108801, AST-2108493, AST-2108569 and the NASA Astrobiology Institute (NNA09DA76A) in the pursuit of precision radial velocities in the NIR. The HPF team was also supported by the Heising-Simons Foundation via grant 2017-0494.

This project was also supported by NSF grant AST-1909682.


Based on observations at Kitt Peak National Observatory, NSF’s NOIRLab (Prop. ID 2020B-0422; PI: A. Lin. Prop. ID 2021A-0385; PI: A. Lin. Prop. ID 2021B-0435; PI: S. Kanodia. Prop ID 2021B-0035; PI: S. Kanodia), managed by the Association of Universities for Research in Astronomy (AURA) under a cooperative agreement with the National Science Foundation. The authors are honored to be permitted
to conduct astronomical research on Iolkam Du’ag (Kitt
Peak), a mountain with particular significance to the
Tohono O’odham.


This paper contains data taken with the NEID instrument, which was funded by the NASA-NSF Exoplanet Observational Research (NN-EXPLORE) partnership and built by Pennsylvania State University. NEID is installed on the WIYN telescope, which is operated by the NSF's National Optical-Infrared Astronomy Research Laboratory (NOIRLab), and the NEID archive is operated by the NASA Exoplanet Science Institute at the California Institute of Technology. NN-EXPLORE is managed by the Jet Propulsion Laboratory, California Institute of Technology under contract with the National Aeronautics and Space Administration.


We thank the NEID Queue Observers and WIYN Observing Associates for their skillful execution of our NEID observations.


This work was partially supported by funding from the Center for Exoplanets and Habitable Worlds.
The Center for Exoplanets and Habitable Worlds is supported by the Pennsylvania State University and the Eberly College of Science.

This research was supported in part by a Seed Grant award from the Institute for Computational and Data Sciences at the Pennsylvania State University.

This work has made use of data from the European Space Agency (ESA) mission {\it Gaia} (\url{https://www.cosmos.esa.int/gaia}), processed by the {\it Gaia} Data Processing and Analysis Consortium (DPAC, \url{https://www.cosmos.esa.int/web/gaia/dpac/consortium}). Funding for the DPAC has been provided by national institutions, in particular the institutions participating in the {\it Gaia} Multilateral Agreement.

This research made use of \textsf{exoplanet} \citep{exoplanet:joss, exoplanet:zenodo} and its
dependencies \citep{celerite1, celerite2, exoplanet:agol20, kumar19, exoplanet:astropy13, exoplanet:astropy18, exoplanet:kipping13, exoplanet:luger18, exoplanet:pymc3, exoplanet:theano}.


This research made use of Lightkurve, a Python package for Kepler and TESS data analysis (Lightkurve Collaboration, 2018).


This research has made use of the SIMBAD database,
operated at CDS, Strasbourg, France

This research has made use of the NASA Exoplanet Archive, which is operated by the California Institute of Technology, under contract with the National Aeronautics and Space Administration under the Exoplanet Exploration Program.


This research has made use of the Exoplanet Follow-up Observation Program (ExoFOP; DOI: 10.26134/ExoFOP5) website, which is operated by the California Institute of Technology, under contract with the National Aeronautics and Space Administration under the Exoplanet Exploration Program.

This research was, in part, carried out at the Jet Propulsion Laboratory, California Institute of Technology, under a contract with the National Aeronautics and Space Administration (80NM0018D0004).

CIC acknowledges support by NASA Headquarters under the NASA Earth and Space Science Fellowship Program through grant 80NSSC18K1114.

\facilities{\gaia{}, HET (HPF), TESS, RBO, APO (ARCTIC), WIYN (NEID), Shane (ShARCS), Exoplanet Archive}
\software{
\texttt{ArviZ} \citep{kumar19}, 
AstroImageJ \citep{collins17}, 
\texttt{astropy} \citep{astropy18},
\texttt{barycorrpy} \citep{kanodia18b}, 
\texttt{exoplanet} \citep{exoplanet:zenodo},
\texttt{ipython} \citep{ipython07},
 \texttt{lightkurve} \citep{lightkurve},
\texttt{matplotlib} \citep{Hunter07},
\texttt{numpy} \citep{harris20},
\texttt{pandas} \citep{reback2020pandas,mckinney-proc-scipy-2010},
\texttt{PyMC3}\citep{exoplanet:pymc3},
\texttt{RadVel}\citep{fulton18},
\texttt{scipy} \citep{2020SciPy-NMeth},
\texttt{SERVAL} \citep{zechmeister18},
\texttt{starry} \citep{exoplanet:luger18},
\texttt{Theano} \citep{exoplanet:theano},
\texttt{TransitLeastSquares} \citep{hippke19}.
}

\bibliography{bibliography}

\appendix

\begin{figure*}[] 
\centering
\includegraphics[width=\textwidth]{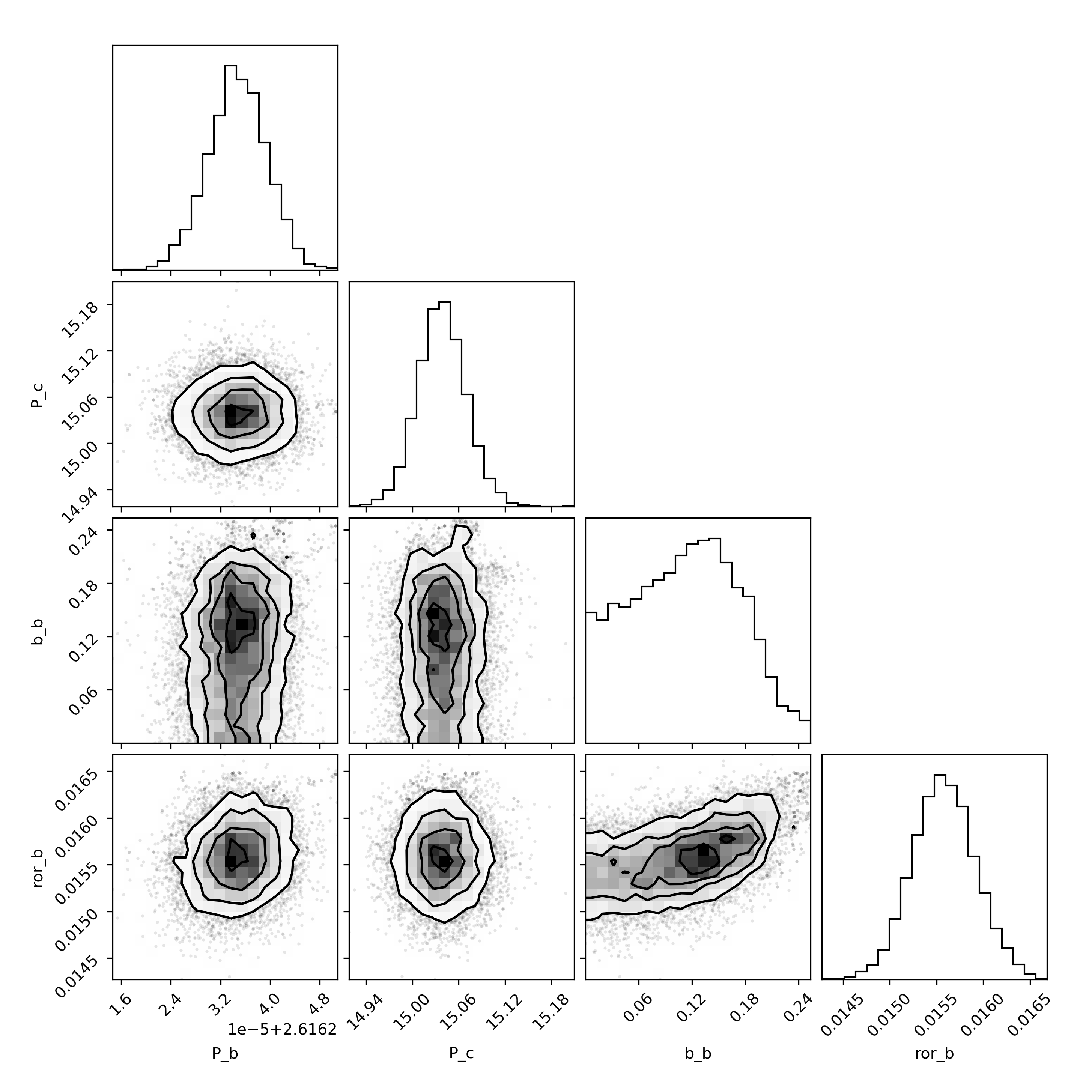}
\caption{We include a corner plot of a few key parameters generated during our joint fit. At the top of each column is a histogram of each parameter's values during the MCMC process, marginalized over other parameters.} \label{fig:corner}
\end{figure*}

We include the RV data take by our team (Tables \ref{tab:neidrv}, \ref{tab:hpfrv}). We include a corner plot of a few of our model parameters in Figure \ref{fig:corner}.

\begin{deluxetable*}{cccccc}
\tablecaption{NEID RVs of GJ 3929. We do not include the five NEID RVs with failed drift solutions. \label{tab:neidrv}}
\tablehead{\colhead{$\unit{BJD_{TDB}}$ (days)}  &  \colhead{RV (m s$^{-1}$)}   & \colhead{$\sigma$ (m s$^{-1}$)} & \colhead{SNR$_{102}$} & \colhead{H$\alpha$ Index} & \colhead{$\sigma$ H$\alpha$ Index}}
\startdata
2459221.0168 & 7.4 & 1.8 & 29.41 & 0.971 & 0.007 \\
2459221.0275 & 1.66 & 1.63 & 32.04 & 0.973 & 0.006 \\
2459221.0384 & 1.05 & 1.69 & 31.09 & 0.965 & 0.006 \\
2459231.0107 & 4.72 & 10.47 & 5.51 & 1.049 & 0.046 \\
2459231.0167 & 8.66 & 1.16 & 43.58 & 0.956 & 0.004 \\
2459231.0274 & 6.81 & 1.29 & 40.16 & 0.956 & 0.005 \\
2459231.0382 & 8.31 & 1.09 & 46.21 & 0.956 & 0.004 \\
2459322.8407 & 11.47 & 1.24 & 43.41 & 0.939 & 0.004 \\
2459327.8212 & 7.42 & 1.08 & 49.62 & 0.941 & 0.004 \\
2459363.9174 & 7.05 & 1.01 & 51.56 & 0.946 & 0.004 \\
2459384.7824 & 12.5 & 1.23 & 42.22 & 0.947 & 0.004 \\
2459385.7494 & 7.97 & 1.48 & 36.59 & 0.957 & 0.005 \\
2459411.8444 & 2.45 & 1.18 & 44.79 & 0.944 & 0.004 \\
2459413.763 & 6.47 & 1.03 & 48.56 & 0.926 & 0.004 \\
2459422.8348 & 3.12 & 1.09 & 47.69 & 0.935 & 0.004 \\
2459424.7787 & 3.12 & 1.9 & 27.83 & 0.93 & 0.007 \\
2459431.7429 & 8.28 & 1.0 & 50.77 & 0.93 & 0.003 \\
2459434.7056 & -0.66 & 1.25 & 41.15 & 0.942 & 0.004 \\
2459475.6962 & 8.87 & 1.33 & 37.61 & 0.965 & 0.005 \\
2459478.6923 & 2.94 & 1.02 & 48.36 & 0.954 & 0.004 \\
2459479.6223 & -0.1 & 0.85 & 54.55 & 0.947 & 0.003 \\
2459481.6797 & 6.08 & 1.57 & 31.97 & 0.957 & 0.006 \\
2459498.5889 & 1.54 & 0.84 & 56.35 & 0.953 & 0.003 \\
2459590.0301 & 6.89 & 0.86 & 54.54 & 0.945 & 0.003 \\
\enddata
\end{deluxetable*}

\begin{deluxetable*}{cccc}
\tablecaption{HPF RVs of GJ 3929. \label{tab:hpfrv}}
\tablehead{\colhead{$\unit{BJD_{TDB}}$ (days)}  &  \colhead{RV (m s$^{-1}$)}   & \colhead{$\sigma$ (m s$^{-1}$)} & \colhead{SNR$_{18}$}}
\startdata
2459088.6272 & -16.43 & 4.98 & 210.68  \\
2459088.6349 & -5.87 & 4.84 & 214.31  \\
2459088.6426 & -3.77 & 5.17 & 202.62  \\
2459222.0314 & -3.74 & 5.48 & 189.58  \\
2459222.0392 & 1.12 & 5.03 & 206.82  \\
2459222.0470 & -4.42 & 5.05 & 207.29  \\
2459233.0037 & 1.42 & 7.62 & 135.48  \\
2459233.0117 & 6.38 & 7.79 & 133.86  \\
2459233.0195 & 12.04 & 7.54 & 140.42  \\
2459271.8998 & 20.83 & 6.97 & 151.17  \\
2459271.9076 & 4.8 & 6.51 & 162.75  \\
2459271.9154 & 3.63 & 7.21 & 146.2  \\
2459296.8246 & 9.54 & 9.38 & 114.91  \\
2459296.8327 & 0.5 & 7.96 & 136.29  \\
2459296.8403 & 6.47 & 7.1 & 147.33  \\
2459649.8572 & -14.68 & 13.81 & 82.09  \\
2459649.8652 & 8.43 & 13.75 & 83.03  \\
2459649.8719 & 4.25 & 25.37 & 49.05  \\
\enddata
\end{deluxetable*}

\end{document}